\documentclass[authoryear,10pt]{article}

\usepackage{amssymb,multirow,amsmath,graphicx,array,lscape,rotating}

\usepackage{graphicx}
\usepackage{subcaption}
\usepackage{enumitem}
\usepackage{xcolor}

\usepackage{PRIMEarxiv}
\usepackage{geometry}
\usepackage[
bibstyle = apa,
citestyle = bwl-FU,
maxcitenames=2,
maxbibnames=99,
natbib=true,backend=biber]{biblatex}
\usepackage[utf8]{inputenc} 
\usepackage[T1]{fontenc}    
\usepackage[hidelinks]{hyperref}       
\usepackage{url}            
\usepackage{booktabs}       
\usepackage{amsfonts}       
\usepackage{nicefrac}       
\usepackage{microtype}      
\usepackage{lipsum}
\usepackage{fancyhdr}       
\usepackage{graphicx}       
\graphicspath{{media/}}     
\usepackage{amssymb}
\usepackage{amsthm}
\usepackage{amsmath}
\usepackage{tabularx}
\usepackage{booktabs}
\usepackage{multicol}
\usepackage{relsize}
\usepackage{float}
\usepackage{xurl}
\usepackage{hyperref}
\usepackage{graphicx}

\usepackage{setspace}

\usepackage{bm}
\usepackage{bbm}

\newcommand{\Unovet}{\bm{1}}

\newcommand{\bvet}{\bm{b}}

\newcommand{\vvet}{\bm{v}}
\newcommand{\wvet}{\bm{w}}
\newcommand{\xvet}{\bm{x}}
\newcommand{\yvet}{\bm{y}}
\newcommand{\zvet}{\bm{z}}
\newcommand{\Avet}{\bm{A}}
\newcommand{\Bvet}{\bm{B}}
\newcommand{\Cvet}{\bm{C}}
\newcommand{\Dvet}{\bm{D}}

\newcommand{\Gvet}{\bm{G}}

\newcommand{\Ivet}{\bm{I}}

\newcommand{\Mvet}{\bm{M}}

\newcommand{\Rvet}{\bm{R}}
\newcommand{\Svet}{\bm{S}}

\newcommand{\Wvet}{\bm{W}}

\newcommand{\Yvet}{\bm{Y}}
\newcommand{\Zvet}{\bm{Z}}
\newcommand{\Zerovet}{\bm{0}}
\newcommand{\Omegavet}{\bm{\Omega}}

\title{Balancing Accuracy and Costs in Cross-Temporal Hierarchies: Investigating Decision-Based and Validation-Based Reconciliation}

\usepackage[affil-it]{authblk}
\author[1,*]{Mahdi Abolghasemi}
\author[2]{Daniele Girolimetto}
\author[3]{Tommaso Di Fonzo}
\affil[1]{School of Mathematical Sciences, Queensland University of Technology, Brisbane, Australia, email: mahdi.abolghasemi@qut.edu.au}
\affil[2]{Department of Statistical Sciences, University of Padova, Padova, Italy, email:daniele.girolimetto@unipd.it }
\affil[2]{Department of Statistical Sciences, University of Padova, Padova, Italy, email: difonzo@stat.unipd.it}

\begin{document}
\maketitle

\begin{abstract}
Wind power forecasting is essential for managing daily operations at wind farms and enabling market operators to manage power uncertainty effectively in demand planning. This paper explores advanced cross-temporal forecasting models and their potential to enhance forecasting accuracy. First, we propose a novel approach that leverages validation errors, rather than traditional in-sample errors, for covariance matrix estimation and forecast reconciliation. Second, we introduce decision-based aggregation levels for forecasting and reconciliation where certain horizons are based on the required decisions in practice. Third, we evaluate the forecasting performance of the models not only on their ability to minimize errors but also on their effectiveness in reducing decision costs, such as penalties in ancillary services. Our results show that statistical-based hierarchies tend to adopt less conservative forecasts and reduce revenue losses. On the other hand, decision-based reconciliation offers a more balanced compromise between accuracy and decision cost, making them attractive for practical use.
\end{abstract}

\keywords{Forecasting \and Decision-Making \and Wind power forecasting  \and Cross-temporal reconciliation}

\section{Introduction}
\label{sec:sample1}

Wind power is a critical resource for the energy market, accounting for over 7.5\% of current electricity generation and projected to grow to 12.1\% of the total market by 2028 \citep{iea}. However, wind power is inherently uncertain and depends on weather conditions, making it intermittent and an unreliable source of generation. As such, accurate wind power forecasting is indispensable in energy markets to effectively integrate wind farms into grid networks.

Modern wind power forecasting methods aim to utilize information across turbines, wind farms, and temporal patterns at various frequencies to improve forecasting accuracy. Hierarchical forecasting models have gained momentum as effective methods due to their ability to leverage cross-sectional, temporal, and cross-temporal information \citep{Hyndman2011,Athanasopoulos2017,DiFonzoGiro2021}. However, constructing effective hierarchies presents significant challenges, particularly in the context of cross-temporal wind power forecasting. Issues such as the stochasticity and intermittency of wind data, along with the large number of time series in a cross-temporal hierarchy, can complicate the estimation of the covariance matrix required for forecast reconciliation \citep{DiFonzoGiro2021,Athanasopoulos2024,english2024improving}. 

Existing approaches typically rely on in-sample forecast errors for estimating the covariance matrix, assuming that similar covariance patterns will hold for future observations. However, this assumption may not be realistic for wind power, given its high variability and stochastic nature. To address this, we propose using validation errors instead of in-sample errors to estimate the covariance matrix. Validation errors provide a more reliable basis for estimation and enhance the robustness of reconciliation in hierarchical forecasting frameworks. Additionally, we leverage high-frequency wind data to effectively build and estimate the covariance matrix, which can otherwise be problematic in cross-temporal hierarchies with low-frequency data and limited observations.

Another important consideration in this study is the efficacy of hierarchical forecasting models for decision-making. Inspired by practical applications, we construct smaller hierarchies for certain temporal frequencies to avoid the computational expense of large cross-temporal hierarchies that may not always be necessary for real-world decision-making \citep{athanasopoulos2023evaluation}. These decision-driven hierarchies are compared with statistical approaches that rely on full cross-temporal models to improve accuracy. We empirically evaluate the trade-offs between decision-based and statistically driven models, examining their performance in practical and operational contexts.

Hierarchical forecasting models are often advocated as great tools for consistent decision-making as they ensure forecasts remain coherent across different levels, enabling seamless decisions. However, it is not always evident whether such forecasts directly lead to consistent decisions or lower costs \citep{athanasopoulos2023evaluation}. This research evaluates the efficacy of forecasting models based on their potential to minimize penalty costs for wind farms. In many electricity markets, wind farms must submit generation forecasts, and their compensation is tied to both their level of generation and their forecast accuracy. Errors in forecasting can result in significant penalties, potentially costing operators millions of dollars \citep{aemo}. Very short-term and short-term forecasts are particularly critical for managing daily wind farm operations and helping market operators address power uncertainty in demand planning. This complex relationship between forecast accuracy and downstream costs underscores the necessity of developing forecasting models that not only enhance accuracy but also reduce operational and financial costs \citep{abolghasemi2023intersection}.

This paper makes three key contributions to hierarchical forecasting for wind power. First, we propose a novel approach that uses validation errors, rather than traditional in-sample errors, for covariance matrix estimation and forecast reconciliation. While in-sample errors are widely used, they may lead to biased estimates when historical data differ significantly from future observations. By leveraging out-of-sample validation errors, we aim to improve the robustness of reconciliation in hierarchical forecasting.

Second, we propose decision-based hierarchies as smaller cross-temporal hierarchies suitable for decision making and investigate their performance against statistical-based hierarchies. Current methods focus on building full cross-temporal hierarchies by exploring all temporal frequencies and cross-sectional aggregations, although some may not be useful for decision-making in practice. We propose an alternative approach, using decision-based aggregation levels where temporal frequencies are selected based on practical decision requirements. This approach reduces unnecessary complexity while maintaining relevance for real-world applications.

Third, we move beyond conventional metrics of forecast accuracy to incorporate decision costs into the evaluation of hierarchical forecasting models. While accuracy is important, the ultimate value of forecasts lies in their ability to support operational and financial decision-making. We assess models not only on their ability to minimize errors but also on their potential to reduce penalties for forecast errors and inefficiencies in grid management. This dual focus ensures the models align with real-world priorities in energy markets.

The remainder of this article is organized as follows. Section \ref{sec:background} reviews relevant studies on wind power forecasting and hierarchical forecasting. Section \ref{sec:methods} provides an overview of common hierarchical forecasting methods. Section \ref{sec:data} describes the experimental setup, including data, pre-processing, base forecasting models, and evaluation methods. Section \ref{sec:results} presents the empirical results and discussions. Finally, Section \ref{sec:conclusion} concludes the paper with key insights and recommendations.

\section{Related literature}
\label{sec:background}

Wind power data are characterized by high frequency, intermittency, multiple seasonality, and unusual ramping on and off behaviour under certain weather conditions, making forecasting wind power a challenging task.  Several methods have been proposed for wind power forecasting, including statistical approaches \citep{lydia2016linear}, physical models \citep{lange2006physical}, and machine learning techniques \citep{wu2021ultra}. Forecasts are often required for very short- to short-term horizons, from a few minutes to a few hours ahead, to be used for energy demand planning. Moreover, forecasts are often needed across different locations and levels to be used for decision-making at regions, or geographical locations \citep{english2024improving}.

Traditional forecasting methods often rely on single time series data, using historical observations of total power generation to predict future outputs. Advances in hierarchical forecasting have introduced new temporal, cross-sectional, and cross-temporal models tailored for wind power forecasting \citep{Gilbert2018,Bai-Pinson2019,Zhang-Dong2019,Sorensen2023EMSM,Sorensen2023WIREs,sharma2024optimal,english2024improving,Zhang2024}. Temporal hierarchies utilize time series data at multiple frequencies, revealing distinct patterns that are useful for different forecasting horizons \citep{sharma2024optimal}. Cross-sectional hierarchies leverage spatial information by integrating data across turbines or locations, improving accuracy through shared insights \citep{english2024improving}. When combined, temporal and cross-sectional hierarchies form cross-temporal hierarchies, which generate forecasts across various locations, levels, and temporal frequencies. These models effectively capture both temporal and spatial information, allowing forecasts to span different turbines, decision-making levels, and time horizons.

Hierarchical forecasts enable decision-makers to monitor power generation across multiple levels, locations, and time horizons. Basic approaches like bottom-up and top-down forecasting achieve forecast consistency by generating forecasts at the bottom or top level of the hierarchy, respectively. However, these approaches fail to utilize the full informational depth of the hierarchy and often underperform compared to more sophisticated reconciliation methods \citep{Hyndman2011}. Sophisticated methods generate forecasts independently for all series in the hierarchy and then reconcile them linearly. Reconciliation for temporal, cross-sectional, and cross-temporal hierarchies requires the estimation of the covariance matrix of base forecast errors. This matrix is then used to linearly adjust base forecasts, ensuring coherent reconciled forecasts.

Accurately estimating the covariance matrix in hierarchical forecasting—particularly for cross-temporal hierarchies—presents a significant challenge. As the number of series increases, the covariance matrix's dimensionality grows, often exceeding the number of observations, which creates technical challenges for computing the covariance matrix \citep{DiFonzoGiro2021}. Various methods have been proposed for estimating covariance matrices in cross-sectional and temporal hierarchies \citep{Hyndman2011,Wickramasuriya2019,sharma2024optimal,Nystrup2020}.

Both cross-sectional and temporal hierarchies share many common methods for reconciliation, but there are nuanced differences between them that require different modeling approaches, as will be discussed. Ordinary least squares (OLS) is the simplest approach for estimating the forecast error covariance matrix, as it ignores the scale of the series and assumes identical error variances and uncorrelated base forecast errors across nodes. However, this rarely holds in practice, and it often lacks optimality due to its inability to incorporate auto- and cross-correlations.

Variance scaling addresses the issues with OLS by scaling the base forecasts using the variance of the residuals, only accounting for the diagonal series while ignoring the others. This approach, also known as the Weighted Least Squares (WLS) method, can improve upon OLS by assigning non-identical error variances. However, it still ignores the off-diagonal elements of the covariance matrix, leaving correlations among nodes unaddressed. Structural scaling is another method that accounts for the hierarchical structure of data but continues to ignore the scale of data and fails to capture the full covariance structure.

Estimating the complete covariance matrix, which includes both variances and covariances, is essential for accommodating correlations across nodes. The Trace Minimization (MinT) method was developed to address this need \citep{Wickramasuriya2019}. MinT integrates the full covariance structure, including off-diagonal elements representing cross-series covariances, resulting in unbiased forecasts with minimum variance. Similarly, other models have been proposed for estimating the covariance matrix in temporal hierarchies to effectively model temporal correlations. For example, \citet{lemos2021probabilistic} modeled correlations using a multivariate predictive covariance of auto-regressive models in a probabilistic load forecasting model for household energy consumption. \citet{Nystrup2020} proposed new estimators that account for both auto-correlation and cross-correlation in reconciling hierarchies. \citet{sharma2024optimal} further extended these models by using the MinT (shrinkage) method to effectively model auto- and cross-correlations.

Estimating the full covariance matrix can be challenging in hierarchical settings, as the number of series grows rapidly. Shrinkage methods have been proposed as an effective solution for estimating this matrix \citep{zhang2005statistical}. These methods rely on in-sample errors and assume they can be used to estimate the covariance matrix for out-of-sample errors. However, this assumption has been challenged in some studies, which argue that it can introduce bias, particularly if future observations differ significantly from historical ones \citep{abolghasemi2022model}. An alternative approach is to use out-of-sample errors for more accurate estimation of this matrix. This has been implemented indirectly through validation samples in machine learning models for reconciling cross-sectional \citep{spiliotis2021hierarchical} and cross-temporal forecasts \citep{rombouts2024cross}. However, direct use of out-of-sample errors for covariance matrix estimation remains unexplored. The high-frequency nature of energy data provides sufficient observations for out-of-sample error estimation, which would be impractical for low-frequency data with fewer observations.

Although hierarchical forecasting generally improves forecasting accuracy, this improvement is not consistent across all levels and nodes in a hierarchy \citep{zhang2023optimal,Athanasopoulos2024}. In hierarchical time series forecasting, some time series may provide stronger signals than others, making them more useful for estimating the covariance matrix. \citet{zhang2023optimal} proposed forecast reconciliation with immutable forecasts, where certain nodes in the hierarchy are fixed while others are reconciled, preserving unbiasedness in the hierarchy. This approach is particularly useful when certain nodes have high accuracy and should not be adjusted. \citet{spiliotis2021hierarchical} proposed machine learning models for reconciliation to improve forecast accuracy at specific levels rather than across all nodes. Using the most relevant information for reconciliation and improving forecast accuracy at specific nodes or levels remains an open problem in hierarchical forecasting.

Forecasts are often used as inputs to inform decision-makers, but they are not the final goal. The utility of forecasts—such as optimizing decisions to maximize profit and minimize penalty costs for wind farms—is more relevant to decision-makers. The ultimate goal of forecasting models is to reduce electricity costs using predictions. While it may seem intuitive that higher forecast accuracy leads to lower costs, the relationship between accuracy and cost is neither linear nor symmetric \citep{abolghasemi2023intersection}. Although accuracy metrics differ from cost metrics, there is limited research on the relationship between forecast accuracy and electricity costs for wind farms. Empirical evidence in this area is sparse, not only for wind power forecasting but also for forecasting in general. In this study, we evaluate models based on both forecast accuracy and decision costs, providing a holistic view of their performance.

We also explore decision-based hierarchical forecasting, where temporal aggregations are constructed only for specific temporal frequencies relevant to decision-making. We aim to build smaller cross-temporal hierarchies to reduce the computational complexity of the models and enable efficient covariance matrix estimation, all while maintaining forecast accuracy.
 
\section{Forecasts reconciliation methods}\label{sec:methods}

In this section, we focus specifically on hierarchical or grouped time series where the top- and bottom-level variables are shared \citep{Hyndman2011}. A comprehensive framework for more complex scenarios, where the time series may involve multiple linearly constrained time series, is presented and discussed in \citep{DiFonzoGiro2021} employing a more general notation.

\subsection{Notation}\label{sec:not}

Consider a straightforward example of a two-level cross-sectional hierarchy
, where the top variable $X$ represents the sum of two lower-level series, $W$ and $Z$. Assume that the highest observed frequency for these variables is 10-min, allowing 20-min, 30-min and 60-min (hour)  time series to be derived through simple non-overlapping aggregation of higher frequency data.

Following the temporal forecast reconciliation literature \citep{Athanasopoulos2017, DiFonzoGiro2021}, the superscript $[k]$ denotes the temporal aggregation order for each level of granularity, with $k = 6$ for hourly, $k = 3$ for 30-minute, $k = 2$ for 20-minute, and $k = 1$ for 10-minute intervals. The cross-sectional hierarchy is defined by the aggregation relationship $X = W + Z$, which holds across any temporal aggregation order $k \in \mathcal{K} = \{6, 3, 2, 1\}$, i.e., $x_{\tau}^{[k]} = w_{\tau}^{[k]} + z_{\tau}^{[k]}$, for $\tau = 1, \dots, m/k$ and $m = \max(\mathcal{K})$.
In matrix notation, let $\yvet_{\tau}^{[k]} = \begin{bmatrix} x_{\tau}^{[k]} & w_{\tau}^{[k]} & z_{\tau}^{[k]} \end{bmatrix}’$ represent the $(3 \times 1)$ vector of observations at temporal granularity $k \in \mathcal{K}$ for time $\tau = 1, \dots, m/k$. The cross-sectional aggregation relationships are expressed as follows:
$$
x_{\tau}^{[k]} = \Avet_{cs} \begin{bmatrix} w_{\tau}^{[k]} \\ z_{\tau}^{[k]} \end{bmatrix}, \quad
\yvet_{\tau}^{[k]} = \Svet_{cs} \begin{bmatrix} w_{\tau}^{[k]} \\ z_{\tau}^{[k]} \end{bmatrix}, \quad
\Cvet_{cs} \yvet_{\tau}^{[k]} = 0
$$
where $\Avet_{cs}$ is the cross-sectional aggregation matrix, $\Svet_{cs}$ is the cross-sectional summing matrix, and $\Cvet_{cs}$ represents cross-sectional constraints in homogeneous form, given by:
$$
\Avet_{cs} = \begin{bmatrix} 1 & 1 \end{bmatrix} ,\quad
\Svet_{cs} = \begin{bmatrix} \Avet_{cs} \\ \Ivet_2 \end{bmatrix} =
\begin{bmatrix} 
	1 & 1 \\ 
	1 & 0 \\ 
	0 & 1 
\end{bmatrix}, \quad
\Cvet_{cs} = \begin{bmatrix} \Ivet_1 & -\Avet_{cs} \end{bmatrix} =
\begin{bmatrix} 1 & -1 & -1 \end{bmatrix},
$$
where $\Ivet_l$ represents the identity matrix of order $l$. On the other hand, to capture the full temporal aggregation relationships for a single variable (such as $V \in \{X, W, Z\}$) across different granularities, we use matrices:
$$
\Avet_{te} = \begin{bmatrix} 
	1 & 1 & 1 & 1 & 1 & 1 \\ 
	1 & 1 &1 & 0 & 0 & 0 \\ 
	0 & 0 & 0 &1 & 1 & 1 \\ 
	1 & 1 &0 & 0 & 0 & 0 \\ 
	0 & 0 & 1 &1 & 0 & 0 \\ 
	0 & 0 & 0 &0 & 1 & 1 \end{bmatrix}, \quad
\Svet_{te} = \begin{bmatrix} \Avet_{te} \\ \Ivet_6 \end{bmatrix} = 
\begin{bmatrix} 
	1 & 1 & 1 & 1 & 1 & 1 \\ 
	1 & 1 &1 & 0 & 0 & 0 \\ 
	0 & 0 & 0 &1 & 1 & 1 \\ 
	1 & 1 &0 & 0 & 0 & 0 \\ 
	0 & 0 & 1 &1 & 0 & 0 \\ 
	0 & 0 & 0 &0 & 1 & 1 \\ 
	\multicolumn{6}{c}{\Ivet_6} 
\end{bmatrix}, \quad
\Cvet_{te} = \begin{bmatrix} \Ivet_6 & -\Avet_{te} \end{bmatrix},
$$
where $\Avet_{te}$, $\Svet_{te}$, and $\Cvet_{te}$ represent the temporal aggregation matrix, temporal summing matrix, and temporal constraints matrix, respectively. Therefore, 
$$
\vvet = \Svet_{cs} \vvet^{[1]}, \quad
\Cvet_{cs} \vvet = \Zerovet_{6\times 1}
$$
where $\vvet = \begin{bmatrix} v^{[4]}_1 & v^{[2]}_1 & v^{[2]}_2 & v^{[1]}_1 & v^{[1]}_2 & v^{[1]}_3 & v^{[1]}_4 \end{bmatrix}’$ and $\vvet^{[1]} = \begin{bmatrix} v^{[1]}_1 & v^{[1]}_2 & v^{[1]}_3 & v^{[1]}4 \end{bmatrix}’$ for $v \in \{x, w, z\}$.

To integrate both cross-sectional and temporal aggregation relationships, all nodes of the cross-temporal hierarchy can be represented based on the hourly series $w_{\tau}^{[1]}$ and $z_{\tau}^{[1]}$, $\tau=1, \ldots, 6$, via the \textit{structural} and \textit{zero-constraints} representation, respectively,
\begin{equation}\label{eq:ctstr}
	\yvet = \Svet_{ct}\bvet^{[1]} \quad \text{and} \quad \Cvet_{ct}, 
\end{equation}
where $\yvet = \begin{bmatrix} \xvet’ & \wvet’ & \zvet’ \end{bmatrix}’$ contains data for all variables at each granularity, $\bvet^{[1]} = \begin{bmatrix} \wvet^{[1]\prime} & \zvet^{[1]\prime} \end{bmatrix}’$ is the high-frequency bottom series vector, $\Cvet_{ct}$ is the cross-temporal constraints matrix \citep{DiFonzoGiro2021} and $\Svet_{ct} = \Svet_{te} \otimes \Svet_{cs}$ \citep{Giro2024} is the summing matrix mapping $\bvet^{[1]}$ to $\yvet$ with $\otimes$ indicates the Kronecker product. Expression (\ref{eq:ctstr}) extends the cross-sectional structural representation from \citep{Hyndman2011}, connecting upper-level data in cross-sectional and temporal hierarchies to the high-frequency bottom-level series. In addition to the number of variables in the cross-sectional hierarchy ($n=3$ in this example), the dimension of $\Svet_{ct}$ is influenced by the temporal granularities that we want to consider. In the application presented later, this effect is explored through two temporal hierarchy sets. Given a 10-minute highest observed frequency, we use a statistical-based strategy aggregating up to 8 hours, $\mathcal{K}_{SB}=\{48, 24, 16, 12, 8, 6, 4, 3, 2, 1\}$, and a decision-based strategy aggregating up to 1 hour, $\mathcal{K}_{DB}=\{6, 3, 2, 1\}$.

\subsection{Forecast reconciliation}\label{sec:fr}
Consider forecasting an $n$-dimensional high-frequency hierarchical time series $\big\{\yvet_t^{[1]}\big\}_{t=1}^T$, with a forecast horizon equal to the temporal aggregation order $m$. For a factor $k$ of $m$, we can define several temporally aggregated versions of each component in $\yvet_t^{[1]}$, based on non-overlapping sums of $k$ successive values, each having a seasonal period of $M_k = m/k$. To avoid ragged-edge data, we assume the total number of observations used in the non-overlapping aggregation is a multiple of $m$, and, then, $N = \frac{T}{m}$ represent the number of observations at the lowest frequency.
Define $\mathcal{K}$ as the set of $p$ factors of $m$ in descending order, $\mathcal{K} = \{k_p, k_{p-1}, \ldots, k_2, k_1\}$, with $k_p = m$ and $k_1 = 1$, and let $m^\ast = \sum_{j=1}^p \frac{m}{k_j}$ and $k^\ast = m^\ast - m$.

Following \citet{DiFonzoGiro2021}, let $\Yvet_{N+h} \equiv \Yvet$ denote the $(n \times m^\ast)$ matrix of target forecasts at any temporal granularity, with a low-frequency forecast horizon $h$, given by:
\[
\Yvet = \left[ \Yvet^{[m]} \; \Yvet^{[k_{p-1}]} \ldots \Yvet^{[k_2]} \; \Yvet^{[1]} \right] =
\left[\begin{array}{c} \Avet \\ \Bvet \end{array}\right] =
\left[\begin{array}{ccccc}
	\Avet^{[m]} & \Avet^{[k_{p-1}]} & \ldots & \Avet^{[k_2]} & \Avet^{[1]} \\
	\Bvet^{[m]} & \Bvet^{[k_{p-1}]} & \ldots & \Bvet^{[k_2]} & \Bvet^{[1]}
\end{array} \right],
\]
where $m$ is the highest available sampling frequency per seasonal cycle. Each matrix $\Yvet^{[k]} = \left[\begin{array}{c} \Avet^{[k]} \ \Bvet^{[k]} \end{array}\right]$ for $k \in \mathcal{K}$ contains the order-$k$ temporal aggregates of the $n_a$ upper-level cross-sectional series $(\Avet^{[k]})$ and the $n_b$ bottom-level cross-sectional series $(\Bvet^{[k]})$, with $n = n_a + n_b$.

Define the matrix of base forecasts $\widehat{\Yvet}$ as:
\[
\widehat{\Yvet} = \left[ \widehat{\Yvet}^{[m]} \; \widehat{\Yvet}^{[k_{p-1}]} \ldots \widehat{\Yvet}^{[k_2]} \; \widehat{\Yvet}^{[1]} \right] =
\left[\begin{array}{ccccc}
	\widehat{\Avet}^{[m]} & \widehat{\Avet}^{[k_{p-1}]} & \ldots & \widehat{\Avet}^{[k_2]} & \widehat{\Avet}^{[1]} \\
	\widehat{\Bvet}^{[m]} & \widehat{\Bvet}^{[k_{p-1}]} & \ldots & \widehat{\Bvet}^{[k_2]} & \widehat{\Bvet}^{[1]}
\end{array}\right].
\]
While target forecasts are expected to be consistent across time and space,
\[
\Cvet_{cs} \Yvet = \Zerovet_{n_a \times m^\ast} \quad \text{and} \quad
\Cvet_{te} \Yvet^{\prime} = \Zerovet_{k^\ast \times n},
\]
base forecasts are generally cross-sectionally and/or temporally inconsistent:
\[
\Cvet_{cs} \widehat{\Yvet} \ne \Zerovet_{n_a \times m^\ast} \quad \text{and/or} \quad
\Cvet_{te} \widehat{\Yvet}^{\prime} \ne \Zerovet_{k^\ast \times n},
\]
where $\Cvet_{cs}$ and $\Cvet_{te}$ are the constraint matrices corresponding to the cross-sectional and temporal frameworks, respectively (see Section~\ref{sec:not}).

The optimal cross-temporal forecast reconciliation solution proposed by \citep{DiFonzoGiro2021} using the projection approach \citep{Stone1942, Byron1978, Byron1979} is given by
\begin{equation}
	\label{eq:oct}
	\widetilde{\yvet} = \left[\Ivet_{nm^\ast} - \Omegavet_{ct}\Cvet’ \left(\Cvet \Omegavet_{ct} \Cvet’\right)^{-1} \Cvet\right] \widehat{\yvet} = \Mvet_{ct} \widehat{\yvet},
\end{equation}
where $\Omegavet_{ct}$ is a positive definite (covariance) matrix and $\Mvet_{ct} =  \Ivet_{nm^\ast} - \Omegavet_{ct}\Cvet’ \left(\Cvet \Omegavet_{ct} \Cvet’\right)^{-1} \Cvet$. 
Alternatively, the cross-temporally reconciled forecasts can be derived \citep{Giro2024} based on the structural approach for cross-sectional reconciliation in \citep{Hyndman2011}:
\begin{equation}
	\label{eq:oct_str}
	\widetilde{\yvet} =  \Svet_{ct} \left(\Svet_{ct}' \Omegavet_{ct}^{-1} \Svet_{ct}\right)^{-1} \Svet_{ct}'\Omegavet_{ct}^{-1} \widehat{\yvet}  = \Svet_{ct}\Gvet_{ct} \widehat{\yvet},
\end{equation}
where $\Gvet_{ct} = \left(\Svet_{ct}' \Omegavet_{ct}^{-1} \Svet_{ct}\right)^{-1} \Svet_{ct}'\Omegavet_{ct}^{-1}$ and $ \Mvet_{ct} =  \Svet_{ct}\Gvet_{ct}$.
Since $\Omegavet_{ct}$ is generally unknown, 
\cite{DiFonzoGiro2021} consider the following approximations for the cross-temporal covariance matrix:
\begin{itemize}[nosep, leftmargin=!, labelwidth=1.75cm, align=right]
	\item[ct$(ols)$:] identity matrix, $\Omegavet_{ct} = \Ivet_{nm^\ast}$;
	\item[ct$(str)$:] structural matrix, $\Omegavet_{ct} = \mathrm{diag}(\Svet_{ct} \Unovet_{mn_b})$;
	\item[ct$(wlsv)$:] series variance scaling matrix, an extension of the variance scaling approach in \citep{Athanasopoulos2017};
	\item[ct$(acov)$:] auto-covariance scaling matrix, based on the auto-covariance extension in \citep{Nystrup2020};
	\item[ct$(bdshr)$:] block-diagonal shrunk cross-covariance matrix where we assume incorrelation along the temporal dimension and shrunk cross-sectional cross-covariance matrix for each temporal aggregation level $k$.
\end{itemize}

\subsection{Partly bottom-up and iterative cross-temporal reconciliation}
\label{sec:heuite}

The bottom-up method is a well-known and fundamental approach in forecast reconciliation, as seen in works such as \citealp{Dunn1976} and \citealp{Dangerfield1992}. This approach generates forecasts for upper-level series by summing up the base forecasts of the bottom-level series within a hierarchy. When dealing with cross-temporal hierarchies, the bottom-up 
approach is represented as
\begin{equation}
	\label{eq:ctbu}
	\widetilde{\yvet}  = \Svet_{ct} {\bvet}^{[1]} = \Svet_{ct}\Gvet_{\text{ct}(bu)}\widehat{\yvet} ,
\end{equation}
where $\Svet_{ct} = \left(\Svet \otimes \Rvet\right)$ is the cross-temporal summing matrix, $\widehat{\bvet}^{[1]} = \text{vec}\left(\widehat{\Bvet}^{[1]\prime}\right)$ is the vector of base forecasts for the high-frequency bottom time series, $\Gvet_{\text{ct}(bu)} = \Gvet_{\text{cs}(bu)} \otimes \Gvet_{\text{te}(bu)}$ with $\otimes$
denoting the Kronecker product,
$$
\Gvet_{\text{cs}(bu)} = \begin{bmatrix}
	\Zerovet_{n_b \times n_a} & \Unovet_{n_b}
\end{bmatrix} \quad \text{and} \quad \Gvet_{\text{te}(bu)} = \begin{bmatrix}
\Zerovet_{m \times k^\ast} & \Unovet_{m}
\end{bmatrix}.
$$

As demonstrated in \citep{DiFonzoGiro2023SE} and \citep{Giro2024}, the cross-temporal bottom-up reconciliation can be viewed as a two-step sequential reconciliation: cross-sectional (temporal) reconciliation of the high-frequency base forecasts $\widehat{\Yvet}^{[1]}$ followed by temporal (cross-sectiona) bottom up. Let $\Gvet_{cs} = \left(\Svet_{cs}' \Wvet^{-1} \Svet_{cs}\right)^{-1} \Svet_{cs}'\Wvet^{-1}$ the weigthed matrix \citep{Hyndman2011} that cross-sectionally reconcile $\widehat{\Yvet}^{[1]}$ with $\Wvet$ a p.d. matrix, then the partly bottom-up reconciled forecasts are obtained as
$$
\widetilde{\yvet}  = \Svet_{ct}\left(\Gvet_{cs} \otimes \Gvet_{\text{te}(bu)}\right)\widehat{\yvet}.
$$
We refer to these approaches as cs$(rec)+$te$(bu)$, where `$rec$' denotes a generic reconciliation approach in the cross-sectional framework \citep{Wickramasuriya2019}.

\cite{DiFonzoGiro2021} propose an iterative method for cross-temporally reconciled forecasts that alternates reconciliation across one dimension (cross-sectional and temporal), inspired by the first two steps of \citep{Kourentzes2019}. For iteration $j \geq 1$, the approach follows these steps:
\begin{description}[nosep]
	\item[Step 1] Compute the temporally reconciled forecasts, denoted $\widetilde{\Yvet}_{te}^{(j)}$, for each variable $i \in {1, \ldots, n}$ using the cross-sectionally reconciled forecasts from the previous iteration, $\widetilde{\Yvet}_{cs}^{(j-1)}$;
	\item[Step 2] Compute the cross-sectionally reconciled forecasts, $\widetilde{\Yvet}_{cs}^{(j)}$, for each temporal aggregation level based on $\widetilde{\Yvet}_{te}^{(j)}$.
\end{description}
The initial values ($j=0$) for the iterations are set to $\widetilde{\Yvet}_{cs}^{(0)} = \widehat{\Yvet}$. The iterative process ends once the entries in the matrix $\Dvet_{te} = \Zvet’ \widetilde{\Yvet}^{(j)\prime}_{cs}$, which holds all temporal discrepancies, become sufficiently small according to the convergence criterion \citep{DiFonzoGiro2021}. In the description above, \textit{temporal-then-cross-sectional} reconciliation is performed iteratively (ite$(rec_{\text{te}},rec_{\text{cs}})$), otherwise the order may be reversed, getting \textit{cross-sectional-then-temporal} reconciliation.

\vspace*{-1cm}
\section{Experiment setup}\label{sec:data}
\subsection{Data}\label{data_des}
\begin{figure}[H] \centering\includegraphics[scale=.99]{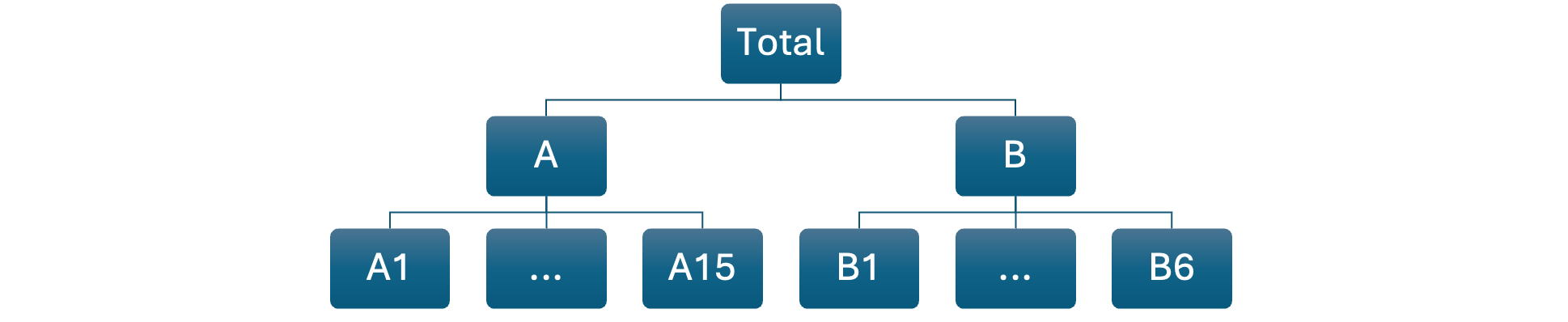} \vspace*{-.1cm} \caption{Cross-sectional hierarchy of turbines}\label{fig:fig1} \end{figure}

We analyzed data from two wind farms in Kelmarsh and Penmanshiel in the UK \citep{plumley_charlie_2022_5841834, plumley_charlie_2022_5946808}, which include a total of 22 wind turbines. The cross-sectional hierarchies of the wind farms are illustrated in Figure \ref{fig:fig1}. The hierarchy comprises three levels: the bottom level represents individual turbines, the middle level aggregates power generation for groups of turbines, and the top level represents the total power generated across all turbines.

\vspace*{-.1cm} 

Each turbine was equipped with a sensor that recorded data at 10-minute intervals, measuring the average wind speed (in m/s) and the average power generated (in kW) during each interval. Some data points were found to be incorrect, typically displaying values of 0 or negative for wind speed or power. These anomalies were attributed to turbine faults, excessively high wind speeds, or other issues. The dataset spans the period from 01/01/2020 to 31/03/2021.

\begin{figure}[H]
\centering\includegraphics[width=0.7\linewidth]{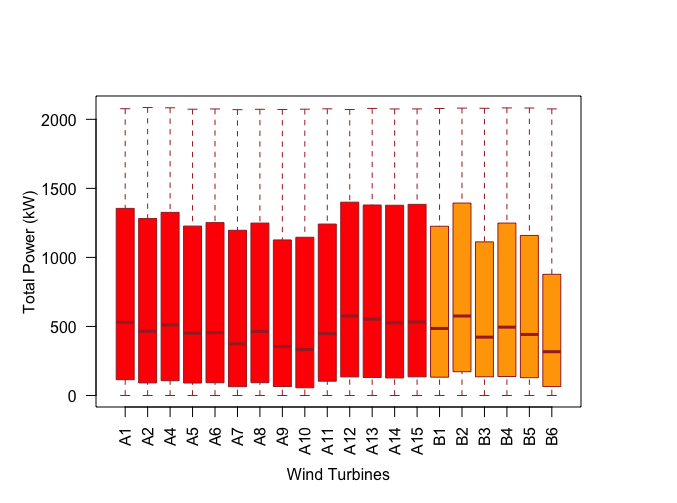}\par
 \vspace*{-.2cm}
\caption{Comparison of 10-minutely power generation for each wind turbine}
 \label{fig:10-min-boxplot}
 \end{figure}

A boxplot of the wind turbines in the hierarchy is presented in Figure \ref{fig:10-min-boxplot}. As shown, turbines of model A have a slightly higher median power generation compared to model B turbines. Turbines of model D, on average, generated greater power than the others, though their median differs from the rest. While the overall power distribution for each turbine group is similar, differences between the groups suggest that separate models may be required to accurately capture their power generation dynamics.

 \vspace*{-.5cm}


\subsection{Preprocessing}\label{subssec3.2}
We pre-processed the data using the following rigorous steps:
i) We checked for negative values of wind power and wind speed and replaced them with 0.
ii) We explored the range and distribution of wind power and wind speeds to ensure they were within reasonable ranges and found no outliers.
iii) We checked for missing values in wind power and wind speed and found none.

After cleaning the data, we extracted useful features and performed extensive feature engineering, which was subsequently used in linear regression and LGBM models. The features are listed in Table \ref{tab:tab1}. Key features included lags of wind speed and wind power, moving averages, moving standard deviations, hour of the day, and seasonal indicators. We considered various feature sets with up to 48 lags for different temporal frequencies. For example, for 10-minutely data, we considered up to 48 lags of wind speed and wind power. For 20-minutely data, we used up to 24 lags, and so on, scaling down to 120-minutely data, where we used up to 4 lags. For lower-frequency data, such as 160, 240, and 480-minutely data, we used up to 3 lags of wind speed and wind power. The same lag lengths were applied when computing moving averages and moving standard deviations of wind speed and wind power. For instance, we calculated moving averages and moving standard deviations over the last 48 steps for 10-minutely data, over the last 24 steps for 20-minutely data, and so on.

To capture the multiple seasonalities present in the data, we employed time-of-day and seasonal dummy variables. Together with moving averages and moving standard deviations, these features effectively captured the behavior of wind power, providing a robust basis for our forecasting models.

\begin{table}[H]\footnotesize
\begin{center}
\caption{Features used for forecasting base time series in regression models}\label{tab:tab1}%
\centerline{\begin{tabular}{@{}lllllllllll@{}}
\toprule
Feature & 10m &20m & 30m& 40m& 60m& 80m& 120m& 160m& 240m& 480m\\
\midrule
Lags of wind speed& 48& 24& 16& 12& 8& 6& 4& 3& 3&3\\
Lags of wind power&48& 24& 16& 12& 8& 6& 4& 3& 3&3\\ 
Wind speed moving average& 1& 1& 1& 1& 1& 1& 1& 1& 1&1\\ 
Wind speed moving standard deviation & 1& 1& 1& 1& 1& 1& 1& 1& 1&1\\
Power moving average& 1& 1& 1& 1& 1& 1& 1& 1& 1 &1\\ 
Power moving standard deviation&  1& 1& 1& 1& 1& 1& 1& 1& 1 &1\\
Quarter& 3& 3& 3& 3& 3& 3& 3& 3& 3&3\\ 
Time of the Day& 23& 23& 23& 23& 23& 19& 11& 9& 3& 2\\
\bottomrule
\end{tabular}}
\end{center}
\end{table}
\unskip

\subsection{Base forecasts}

We employed several forecasting methods to generate the base time series forecasts. These included naive models as a benchmarking method, linear regression as a simple yet effective model, and the LightGBM model, which has demonstrated strong performance in various forecasting tasks, including being the winning method for forecasting hierarchical time series in the M5 competition \citep{MAKRIDAKIS20221346}.  

The linear regression (LR) model was chosen as a baseline due to its simplicity and interpretability. We used the \texttt{fable} package in R to generate the base forecasts for LR models \citep{fable}. Additionally, we utilized the Light Gradient Boosting Machine (LGBM) method as a more advanced forecasting technique. LGBM is a tree-based method that employs a leaf-wise growth strategy and boosting techniques to construct an ensemble of weak learners, resulting in a strong and computationally efficient model \citep{friedman2001greedy, ke2017lightgbm}. We used the \texttt{lightGBM} package in R for training the LGBM models \citep{lightgbm}.  

For both LR and LGBM methods, we trained one model per turbine and temporal frequency, resulting in 23 models for each level of temporal granularity (e.g., 10-, 20-, 30-, 60-, 80-, 120-, 160-, 240-, and 480-minutely data). All forecasts were generated using a rolling origin approach for an 8-hour horizon. The number of forecasting steps depended on the temporal frequency. For example, for 10-minutely data, we generated 48 steps corresponding to 8 hours into the future. For 20-minutely data, we generated 24 steps; for 30-minutely data, 16 steps; and so on. The features listed in Table \ref{tab:tab1} were used as explanatory variables for generating the base forecasts. These $h$-step-ahead forecasts were subsequently reconciled cross-temporally using the methods described in Section \ref{sec:methods}.  

\subsection*{Hyperparameter Optimization}

The hyperparameters for the LGBM models were optimized using a Bayesian optimization search algorithm. We used data from 2020-01-01 to 2020-10-01 for training and data from 2020-10-01 to 2020-12-31 for evaluation. Hyperparameter optimization was conducted on the evaluation set. To reduce computational time, we trained a single model for each temporal frequency and used it to forecast the entire evaluation set during optimization.  

The hyperparameters considered for optimization included, maximum number of leaves in a tree (\textit{num\_leaves}), maximum depth of a tree (\textit{max\_depth}), learning rate (\textit{learning\_rate}), and minimum number of data points in a leaf (\textit{min\_data\_in\_leaf}). We used the following parameter ranges during optimization:  
\begin{itemize}[nosep]
    \item \textit{learning\_rate}: [0.01, 0.1]  
    \item \textit{max\_depth}: [6, 10]  
    \item \textit{num\_leaves}: [50, 200]  
    \item \textit{min\_data\_in\_leaf}: [100, 250]  
\end{itemize}

The optimal hyperparameters, determined through Bayesian optimization, varied across temporal aggregation levels for both validation and in-sample datasets. These results are summarized in Table \ref{tab:hyperparams}. We employed the \texttt{rBayesianOptimization} package in R to perform hyperparameter optimization \citep{rBayes}.

\begin{table}[H]\footnotesize
\begin{center}
\caption{Optimal hyperparameters values for various temporal aggregations for LGBM models \label{tab:hyperparams}}
		\centerline{\begin{tabular}{@{}lllllllllll@{}}
			\toprule
			Approach & Hyperparameters & 10m & 20m & 30m & 60m & 80m & 120m & 160m & 240m& 480m\\
   			\midrule
In-sample 
  &learning\_rate &0.08  &  0.1& 0.1  & 0.1  & 0.08  & 0.09  & 0.09 &0.1& 0.1\\
&max\_depth &  10& 6 &  10&  10 &  6 & 10  & 9&6&6\\
 & num\_leaves & 83& 199  & 50  & 50  &  170& 50  & 149  &198&53 \\
 & min\_data\_in\_leaf &  170 & 156  &  238 &245  & 204  & 111  &140   &250 &100\\
 
\hline
 Validation
  &learning\_rate & 0.09 &0.08  & 0.09 & 0.09 &  0.08 &  0.09 & 0.1  & 0.09&0.1\\
&max\_depth &9  &10  & 7 &  6& 6  & 6  & 6  &8 &6\\
 & num\_leaves &61 & 144  &  96 & 57  & 58 &  50 & 188  &  182 & 50\\
 & min\_data\_in\_leaf &225   &  228 &  100 &  102& 249  & 250  & 240  & 250 & 100\\
			\bottomrule
		\end{tabular}}
\end{center}
\end{table}

\subsection{Forecast reconciliation}

To obtain the reconciled forecasts, we use several approaches available in the R package \texttt{FoReco} \citep{FoReco}, each offering unique advantages in terms of accuracy and computational aspects:
\begin{itemize}[nosep]
\item[pbu] Cross-sectional reconciliation approach using shrinkage estimation of the covariance matrix \citep{Wickramasuriya2019}, alongside a bottom-up aggregation at the temporal level, $\text{cs}(shr)+\text{te}(bu)$ (see the partly bottom-up description in Section~\ref{sec:heuite}).
\item[ct($\,\cdot\,$)] optimal cross-temporal forecast reconciliation approach using four different covariance matrix approximations -- $\text{ct}(str)$, $\text{ct}(wlsv)$, $\text{ct}(bdshr)$, $\text{ct}(acov)$ -- presented in Section~\ref{sec:fr}
\item[ite] iterative approach (see Section~\ref{sec:heuite}) alternating temporal and cross-sectional reconciliation steps, $\text{ite}(acov_{\text{te}}, shr_{\text{cs}})$. Specifically, temporal reconciliation is performed using an auto-covariance scaling ($acov$) matrix, while the cross-sectional reconciliation is achieved through shrinkage ($shr$).
\end{itemize}
To ensure non-negativity, all negative values were set to zero for both the base forecasts' models (lr and lgbm). For the reconciliation process, the \textit{sntz} (set-negative-to-zero) procedure was utilized \citep{DiFonzoGiro2023SE}.


Two distinct strategies have been implemented to construct the temporal hierarchy used in the reconciliation process. The first approach (statistical-based, SB) involves aggregating data across multiple temporal scales, beginning at 10-minute intervals and extending up to 480 minutes (equivalent to 8 hours). This strategy aims to capture a comprehensive view of temporal patterns by leveraging a wide range of aggregation levels, including all possible temporal relationships.
The second approach (decision-based, DB) focuses on a more selective aggregation scheme, using specific intervals of 10, 20, 30, and 60 minutes. This choice is informed by domain-specific requirements or prior analyses suggesting that these intervals offer optimal insights for modeling or decision-making purposes. Unlike the statistical-based strategy, which aims for exhaustive coverage, the decision-based strategy emphasizes targeted aggregation, potentially simplifying computational demands while focusing on the most informative temporal resolutions \citep{athanasopoulos2023evaluation}.

In addition, two different methodologies were employed for the estimation of the covariance matrix, crucial for the accurate reconciliation of the forecasts. The first method is based on the standard literature on forecast reconciliation \citep{Wickramasuriya2019}, which uses in-sample errors to estimate the covariance matrix. In-sample errors, calculated using the training data, offer a straightforward way to model the relationships between forecast errors, benefiting from established statistical properties derived during model fitting.
The second methodology diverges by introducing a validation set of 92 days, used to estimate the covariance matrix based on the “true” forecast errors. This validation-based approach aims to provide a more realistic assessment of forecast performance, potentially leading to improved model robustness by ensuring that the covariance estimation is informed by data not previously seen by the model. 

\section{Empirical results and discussion}\label{sec:results}

\subsection{Forecast accuracy evaluation}

\begin{figure}[!th]
    \centering
    \includegraphics[width=\linewidth]{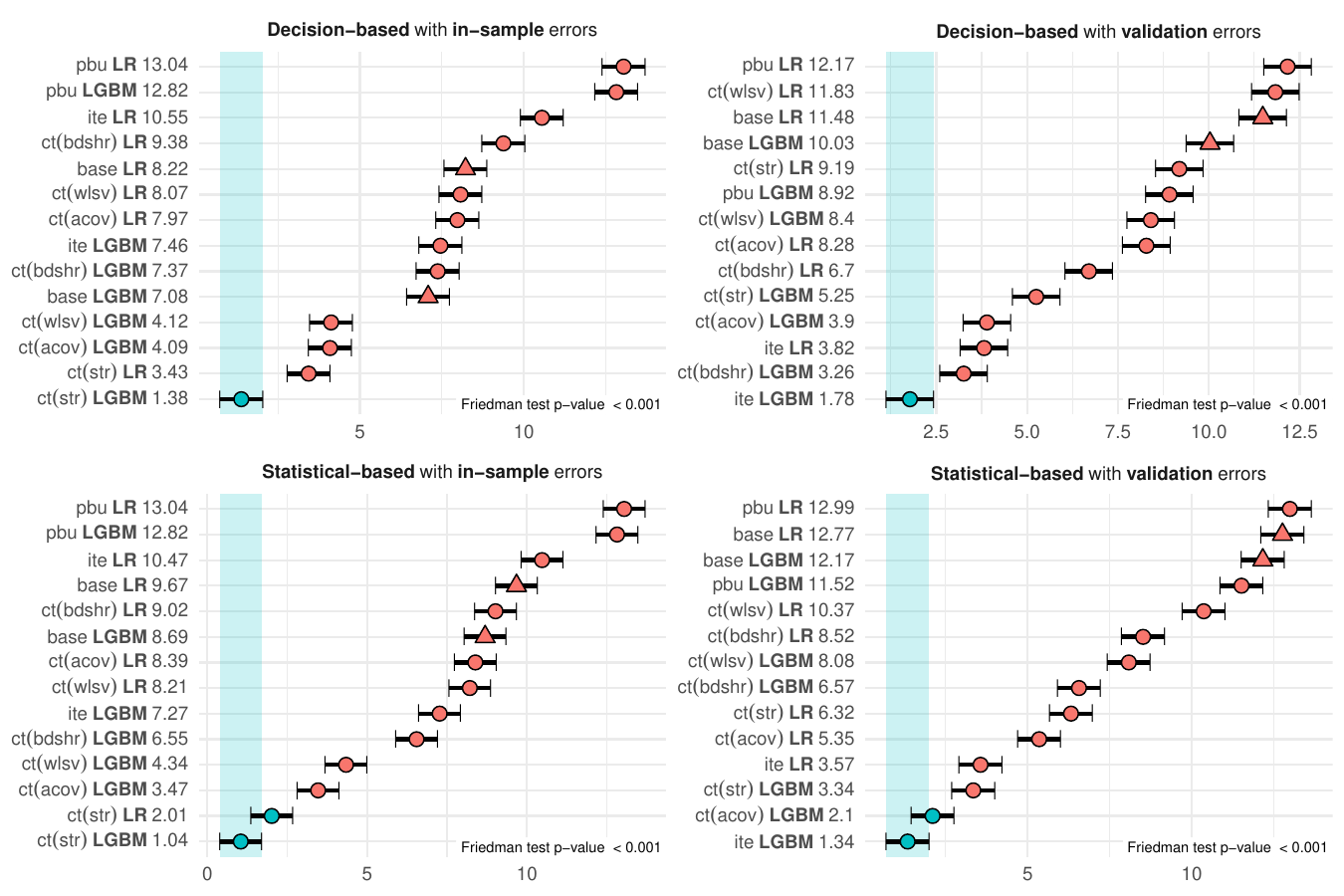}
    \caption{MCB Nemenyi test for all the temporal aggregation levels divided by temporal hierarchy (decision- or statistical-based) and errors (in-sample or validation). In each panel, the Friedman test p-value is reported in the lower-right corner. The mean rank of each approach is shown to the right of its name. Statistically significant differences in performance are indicated if the intervals of two forecast reconciliation procedures do not overlap. Thus, approaches that do not overlap with the green interval are considered significantly worse than the best, and vice versa.}
    \label{fig:mcb}
\end{figure}

To assess the quality of the forecasts, Tables \ref{tab:mse_vl_sb},  \ref{tab:mse_in_sb}, \ref{tab:mse_vl_db}, and \ref{tab:mse_in_db} report the Average Relative Mean Squared Error \citep{Fleming1986, Davydenko2013} for each base model (LR or LGBM), error type (validation or in-sample), and temporal hierarchy (statistical- or decision-based):
$$
AvgRelMSE^{[k]}_j = \left(\prod_{i = 1}^{23} \frac{MSE_{i, j}^{[k]}}{MSE_{i, b}^{[k]}}\right)^{\frac{1}{23}} \quad \text{with} \quad MSE_{i, j}^{[k]} = \frac{1}{QH_{k}}\sum_{q = 1}^Q \sum_{h = 1}^{H_k}\left(\bar{y}_{q, h, i, j}^{[k]}-y_{q, h, i}^{[k]}\right)^2
$$
where $\bar{y}$ represents either the base forecast ($\widehat{y}$) or the reconciled forecast ($\widetilde{y}$), $q = 1, \ldots, Q$ denotes the index of the forecast origins in the test set, $h = 1, \ldots, H_k$ corresponds to the forecast horizon ($H_k = m/k$), $i = 1, \ldots, 23$ indexes the individual time series, $j$ indicates forecast approach, and $b$ refers to the benchmark model (naive). The indices $AvgRelMSE^{[k]}$ provide a normalized measure of error, allowing a consistent assessment relative to the benchmark. Furthermore, in \autoref{fig:mcb} we apply the nonparametric Friedman test and the post hoc multiple comparison with the best (MCB) Nemenyi test \citep{Koning2005, Kourentzes2019, Makridakis2022} to determine whether the forecasting performance of the evaluated techniques differs significantly. 

\begin{table}[p]
    \centering
    
\begin{tabular}[t]{>{}l|ccccccccccl}
\toprule
\multicolumn{1}{c}{\textbf{}} & \multicolumn{11}{c}{\textbf{Temporal aggregation orders}} \\
\cmidrule(l{0pt}r{0pt}){2-12}
\multicolumn{1}{l|}{\textbf{Approach}} & 10 & 20 & 30 & 40 & 60 & 80 & 120 & 160 & 240 & 480 & All\\
\midrule
\addlinespace[0.3em]
\multicolumn{12}{l}{\textbf{LR base models}}\\
base & 0.968 & 0.961 & 0.954 & 0.951 & 0.942 & 0.936 & 0.924 & 0.897 & 0.874 & 0.881 & 0.928\\
$\text{pbu}$ & 0.948 & 0.966 & 0.964 & 0.958 & 0.946 & 0.938 & 0.919 & 0.902 & 0.850 & 0.691 & 0.904\\
$\text{ct}(str)$ & 0.900 & 0.916 & 0.913 & 0.906 & 0.894 & 0.885 & 0.864 & 0.846 & 0.794 & 0.635 & 0.851\\
$\text{ct}(wlsv)$ & 0.927 & 0.944 & 0.942 & 0.935 & 0.923 & 0.915 & 0.895 & 0.877 & 0.825 & 0.666 & 0.881\\
$\text{ct}(bdshr)$ & 0.916 & 0.932 & 0.930 & 0.923 & 0.911 & 0.903 & 0.882 & 0.865 & 0.813 & 0.654 & 0.869\\
$\text{ct}(acov)$ & 0.897 & 0.912 & 0.909 & 0.902 & 0.890 & 0.881 & 0.861 & 0.843 & 0.792 & 0.637 & 0.848\\
$\text{ite}$ & \em{0.890} & \em{0.905} & \em{0.902} & \em{0.895} & \em{0.882} & \em{0.874} & \em{0.854} & \em{0.835} & \em{0.784} & \em{0.629} & \em{0.841}\\
\addlinespace[0.3em]
\multicolumn{12}{l}{\textbf{LGBM base models}}\\
base & 0.962 & 0.952 & 0.947 & 0.940 & 0.932 & 0.925 & 0.917 & 0.908 & 0.896 & 0.945 & 0.932\\
$\text{pbu}$ & 0.937 & 0.955 & 0.953 & 0.946 & 0.934 & 0.926 & 0.907 & 0.889 & 0.838 & 0.679 & 0.892\\
$\text{ct}(str)$ & 0.888 & 0.903 & 0.900 & 0.893 & 0.880 & 0.871 & 0.850 & 0.832 & 0.779 & 0.621 & 0.837\\
$\text{ct}(wlsv)$ & 0.916 & 0.933 & 0.930 & 0.923 & 0.911 & 0.903 & 0.883 & 0.865 & 0.813 & 0.655 & 0.869\\
$\text{ct}(bdshr)$ & 0.904 & 0.920 & 0.917 & 0.910 & 0.898 & 0.889 & 0.869 & 0.851 & 0.799 & 0.641 & 0.855\\
$\text{ct}(acov)$ & 0.882 & 0.897 & 0.894 & 0.887 & 0.874 & 0.865 & 0.845 & 0.827 & 0.776 & 0.622 & 0.832\\
$\text{ite}$ & \em{\textbf{0.875}} & \em{\textbf{0.890}} & \em{\textbf{0.886}} & \em{\textbf{0.879}} & \em{\textbf{0.866}} & \em{\textbf{0.857}} & \em{\textbf{0.837}} & \em{\textbf{0.819}} & \em{\textbf{0.767}} & \em{\textbf{0.613}} & \em{\textbf{0.825}}\\
\bottomrule
\end{tabular}

    \vskip1em
    \caption{$AvgRelMSE$ results are presented using \textbf{validation} errors and a \textbf{statistical-based} temporal hierarchy, categorized by base models (LR or LGBM) and temporal aggregation orders (10, 20, 30, 40, 60, 80, 120, 160, 240, 480, and “All” summarizing results across all orders). The best result for a fixed base model, error type, and temporal hierarchy is indicated in \textit{italic}, the best result for a fixed error type and temporal hierarchy is in \textbf{bold}, and the overall best result is highlighted in \textcolor{blue}{blue}.}
    \label{tab:mse_vl_sb}
\vskip1cm
    \centering
    
\begin{tabular}[t]{>{}l|ccccccccccl}
\toprule
\multicolumn{1}{c}{\textbf{}} & \multicolumn{11}{c}{\textbf{Temporal aggregation orders}} \\
\cmidrule(l{0pt}r{0pt}){2-12}
\multicolumn{1}{l|}{\textbf{Approach}} & 10 & 20 & 30 & 40 & 60 & 80 & 120 & 160 & 240 & 480 & All\\
\midrule
\addlinespace[0.3em]
\multicolumn{12}{l}{\textbf{LR base models}}\\
base & 0.968 & 0.961 & 0.954 & 0.951 & 0.942 & 0.936 & 0.924 & 0.897 & 0.874 & 0.881 & 0.928\\
$\text{pbu}$ & 0.979 & 0.998 & 0.997 & 0.991 & 0.979 & 0.972 & 0.953 & 0.937 & 0.886 & 0.726 & 0.938\\
$\text{ct}(str)$ & \em{0.900} & \em{0.916} & \em{0.913} & \em{0.906} & \em{0.894} & \em{0.885} & \em{0.864} & \em{0.846} & \em{0.794} & \em{0.635} & \em{0.851}\\
$\text{ct}(wlsv)$ & 0.941 & 0.959 & 0.957 & 0.950 & 0.939 & 0.931 & 0.911 & 0.894 & 0.842 & 0.683 & 0.897\\
$\text{ct}(bdshr)$ & 0.944 & 0.962 & 0.960 & 0.953 & 0.942 & 0.934 & 0.914 & 0.897 & 0.846 & 0.686 & 0.900\\
$\text{ct}(acov)$ & 0.942 & 0.959 & 0.957 & 0.950 & 0.939 & 0.931 & 0.911 & 0.894 & 0.842 & 0.683 & 0.897\\
$\text{ite}$ & 0.949 & 0.967 & 0.965 & 0.959 & 0.947 & 0.939 & 0.920 & 0.902 & 0.851 & 0.692 & 0.905\\
\addlinespace[0.3em]
\multicolumn{12}{l}{\textbf{LGBM base models}}\\
base & 0.962 & 0.952 & 0.947 & 0.940 & 0.932 & 0.925 & 0.917 & 0.908 & 0.896 & 0.945 & 0.932\\
$\text{pbu}$ & 0.977 & 0.997 & 0.995 & 0.989 & 0.978 & 0.971 & 0.952 & 0.935 & 0.884 & 0.725 & 0.936\\
$\text{ct}(str)$ & \em{\textbf{0.888}} & \em{\textbf{0.903}} & \em{\textbf{0.900}} & \em{\textbf{0.893}} & \em{\textbf{0.880}} & \em{\textbf{0.871}} & \em{\textbf{0.850}} & \em{\textbf{0.832}} & \em{\textbf{0.779}} & \em{\textbf{0.621}} & \em{\textbf{0.837}}\\
$\text{ct}(wlsv)$ & 0.930 & 0.947 & 0.945 & 0.938 & 0.926 & 0.918 & 0.899 & 0.881 & 0.829 & 0.671 & 0.884\\
$\text{ct}(bdshr)$ & 0.937 & 0.955 & 0.953 & 0.946 & 0.934 & 0.926 & 0.906 & 0.889 & 0.838 & 0.679 & 0.892\\
$\text{ct}(acov)$ & 0.930 & 0.946 & 0.944 & 0.937 & 0.926 & 0.917 & 0.898 & 0.880 & 0.828 & 0.670 & 0.883\\
$\text{ite}$ & 0.939 & 0.956 & 0.954 & 0.947 & 0.935 & 0.927 & 0.908 & 0.890 & 0.839 & 0.680 & 0.893\\
\bottomrule
\end{tabular}

    \vskip1em
    \caption{$AvgRelMSE$ results are presented using \textbf{in-sample} errors and a \textbf{statistical-based} temporal hierarchy, categorized by base models (LR or LGBM) and temporal aggregation orders (10, 20, 30, 40, 60, 80, 120, 160, 240, 480, and “All,” summarizing results across all orders). The best result for a fixed base model, error type, and temporal hierarchy is indicated in \textit{italic}, the best result for a fixed error type and temporal hierarchy is in \textbf{bold}, and the overall best result is highlighted in \textcolor{blue}{blue}.}
    \label{tab:mse_in_sb}
\end{table}

\begin{table}[p]
    \centering
    
\begin{tabular}[t]{>{}l|ccccccccccl}
\toprule
\multicolumn{1}{c}{\textbf{}} & \multicolumn{11}{c}{\textbf{Temporal aggregation orders}} \\
\cmidrule(l{0pt}r{0pt}){2-12}
\multicolumn{1}{l|}{\textbf{Approach}} & 10 & 20 & 30 & 40 & 60 & 80 & 120 & 160 & 240 & 480 & All\\
\midrule
\addlinespace[0.3em]
\multicolumn{12}{l}{\textbf{LR base models}}\\
base & 0.968 & 0.961 & 0.954 & 0.951 & 0.942 & 0.936 & 0.924 & 0.897 & 0.874 & 0.881 & 0.928\\
$\text{pbu}$ & 0.948 & 0.966 & 0.964 & 0.958 & 0.946 & 0.938 & 0.919 & 0.902 & 0.850 & 0.691 & 0.904\\
$\text{ct}(str)$ & 0.937 & 0.955 & 0.953 & 0.946 & 0.934 & 0.926 & 0.906 & 0.889 & 0.837 & 0.678 & 0.892\\
$\text{ct}(wlsv)$ & 0.944 & 0.962 & 0.960 & 0.953 & 0.942 & 0.934 & 0.914 & 0.897 & 0.845 & 0.686 & 0.900\\
$\text{ct}(bdshr)$ & 0.931 & 0.948 & 0.945 & 0.939 & 0.927 & 0.919 & 0.899 & 0.881 & 0.830 & 0.671 & 0.885\\
$\text{ct}(acov)$ & 0.935 & 0.952 & 0.950 & 0.943 & 0.931 & 0.923 & 0.903 & 0.886 & 0.834 & 0.675 & 0.889\\
$\text{ite}$ & \em{0.923} & \em{0.940} & \em{0.938} & \em{0.931} & \em{0.919} & \em{0.910} & \em{0.891} & \em{0.873} & \em{0.821} & \em{0.662} & \em{0.876}\\
\addlinespace[0.3em]
\multicolumn{12}{l}{\textbf{LGBM base models}}\\
base & 0.962 & 0.952 & 0.947 & 0.940 & 0.932 & 0.925 & 0.917 & 0.908 & 0.896 & 0.945 & 0.932\\
$\text{pbu}$ & 0.937 & 0.955 & 0.953 & 0.946 & 0.934 & 0.926 & 0.907 & 0.889 & 0.838 & 0.679 & 0.892\\
$\text{ct}(str)$ & 0.928 & 0.945 & 0.943 & 0.936 & 0.924 & 0.916 & 0.896 & 0.879 & 0.827 & 0.668 & 0.882\\
$\text{ct}(wlsv)$ & 0.935 & 0.953 & 0.951 & 0.944 & 0.932 & 0.924 & 0.904 & 0.887 & 0.835 & 0.677 & 0.890\\
$\text{ct}(bdshr)$ & 0.920 & 0.936 & 0.934 & 0.927 & 0.915 & 0.907 & 0.887 & 0.869 & 0.817 & 0.659 & 0.873\\
$\text{ct}(acov)$ & 0.925 & 0.942 & 0.939 & 0.932 & 0.920 & 0.912 & 0.892 & 0.875 & 0.823 & 0.664 & 0.878\\
$\text{ite}$ & \em{\textbf{0.914}} & \em{\textbf{0.931}} & \em{\textbf{0.928}} & \em{\textbf{0.921}} & \em{\textbf{0.909}} & \em{\textbf{0.900}} & \em{\textbf{0.880}} & \em{\textbf{0.862}} & \em{\textbf{0.810}} & \em{\textbf{0.652}} & \em{\textbf{0.867}}\\
\bottomrule
\end{tabular}

    \vskip1em
    \caption{$AvgRelMSE$ results are presented using \textbf{validation} errors and a \textbf{decision-based} temporal hierarchy, categorized by base models (LR or LGBM) and temporal aggregation orders (10, 20, 30, 40, 60, 80, 120, 160, 240, 480, and “All” summarizing results across all orders). The best result for a fixed base model, error type, and temporal hierarchy is indicated in \textit{italic}, the best result for a fixed error type and temporal hierarchy is in \textbf{bold}, and the overall best result is highlighted in \textcolor{blue}{blue}.}
    \label{tab:mse_vl_db}
\vskip1cm
    \centering
    
\begin{tabular}[t]{>{}l|ccccccccccl}
\toprule
\multicolumn{1}{c}{\textbf{}} & \multicolumn{11}{c}{\textbf{Temporal aggregation orders}} \\
\cmidrule(l{0pt}r{0pt}){2-12}
\multicolumn{1}{l|}{\textbf{Approach}} & 10 & 20 & 30 & 40 & 60 & 80 & 120 & 160 & 240 & 480 & All\\
\midrule
\addlinespace[0.3em]
\multicolumn{12}{l}{\textbf{LR base models}}\\
base & 0.968 & 0.961 & 0.954 & 0.951 & 0.942 & 0.936 & 0.924 & 0.897 & 0.874 & 0.881 & 0.928\\
$\text{pbu}$ & 0.979 & 0.998 & 0.997 & 0.991 & 0.979 & 0.972 & 0.953 & 0.937 & 0.886 & 0.726 & 0.938\\
$\text{ct}(str)$ & \em{0.937} & \em{0.955} & \em{0.953} & \em{0.946} & \em{0.934} & \em{0.926} & \em{0.906} & \em{0.889} & \em{0.837} & \em{0.678} & \em{0.892}\\
$\text{ct}(wlsv)$ & 0.950 & 0.968 & 0.966 & 0.960 & 0.948 & 0.940 & 0.921 & 0.904 & 0.852 & 0.693 & 0.906\\
$\text{ct}(bdshr)$ & 0.956 & 0.974 & 0.972 & 0.966 & 0.954 & 0.947 & 0.927 & 0.910 & 0.859 & 0.700 & 0.912\\
$\text{ct}(acov)$ & 0.950 & 0.968 & 0.966 & 0.960 & 0.948 & 0.940 & 0.921 & 0.904 & 0.852 & 0.693 & 0.906\\
$\text{ite}$ & 0.958 & 0.976 & 0.975 & 0.968 & 0.957 & 0.949 & 0.930 & 0.913 & 0.862 & 0.702 & 0.915\\
\addlinespace[0.3em]
\multicolumn{12}{l}{\textbf{LGBM base models}}\\
base & 0.962 & 0.952 & 0.947 & 0.940 & 0.932 & 0.925 & 0.917 & 0.908 & 0.896 & 0.945 & 0.932\\
$\text{pbu}$ & 0.977 & 0.997 & 0.995 & 0.989 & 0.978 & 0.971 & 0.952 & 0.935 & 0.884 & 0.725 & 0.936\\
$\text{ct}(str)$ & \em{\textbf{0.928}} & \em{\textbf{0.945}} & \em{\textbf{0.943}} & \em{\textbf{0.936}} & \em{\textbf{0.924}} & \em{\textbf{0.916}} & \em{\textbf{0.896}} & \em{\textbf{0.879}} & \em{\textbf{0.827}} & \em{\textbf{0.668}} & \em{\textbf{0.882}}\\
$\text{ct}(wlsv)$ & 0.941 & 0.959 & 0.957 & 0.950 & 0.939 & 0.931 & 0.911 & 0.894 & 0.842 & 0.683 & 0.897\\
$\text{ct}(bdshr)$ & 0.950 & 0.968 & 0.966 & 0.960 & 0.948 & 0.940 & 0.921 & 0.904 & 0.853 & 0.693 & 0.906\\
$\text{ct}(acov)$ & 0.941 & 0.959 & 0.957 & 0.950 & 0.939 & 0.931 & 0.911 & 0.894 & 0.842 & 0.683 & 0.897\\
$\text{ite}$ & 0.950 & 0.968 & 0.967 & 0.960 & 0.948 & 0.941 & 0.921 & 0.904 & 0.853 & 0.694 & 0.907\\
\bottomrule
\end{tabular}

    \vskip1em
    \caption{$AvgRelMSE$ results are presented using \textbf{in-sample} errors and a \textbf{decision-based} temporal hierarchy, categorized by base models (LR or LGBM) and temporal aggregation orders (10, 20, 30, 40, 60, 80, 120, 160, 240, 480, and “All” summarizing results across all orders). The best result for a fixed base model, error type, and temporal hierarchy is indicated in \textit{italic}, the best result for a fixed error type and temporal hierarchy is in \textbf{bold}, and the overall best result is highlighted in \textcolor{blue}{blue}.}
    \label{tab:mse_in_db}
\end{table}

The results of our forecast evaluation consistently demonstrate that all approaches, whether base models or reconciled forecasts, lead to improvements over the naive benchmark. In particular, the reconciliation methods, particularly the optimal and iterative approaches (denoted ct), consistently outperform their corresponding base forecasts across various temporal aggregation levels. These improvements are especially pronounced at higher aggregation levels (e.g. $k \in \{60, 120, 240, 480\}$), where the temporal dependencies captured by reconciliation strategies contribute significantly to the accuracy of the forecast.

However, the partly bottom-up approach (pbu) shows competitive performance when using validation errors. However, when in-sample residuals are used, the pbu approach often performs worse than base models, with exceptions occurring only at higher aggregation levels ($k \in \{240, 480\}$). Moreover, when considering only in-sample errors, the reconciliation approaches fail to outperform the ct($str$) approach, which assumes a diagonal covariance matrix with variances dependent only on the cross-sectional and temporal hierarchy. 

In general, the iterative approach (ite) using validation errors and a statistically based temporal hierarchy emerges as the best-performing strategy in all evaluation scenarios. This method consistently achieves lower AvgRelMSE values, confirming its robustness in reducing forecast errors. 

\begin{figure}[p]
    \centering
    \includegraphics[width=0.9\linewidth]{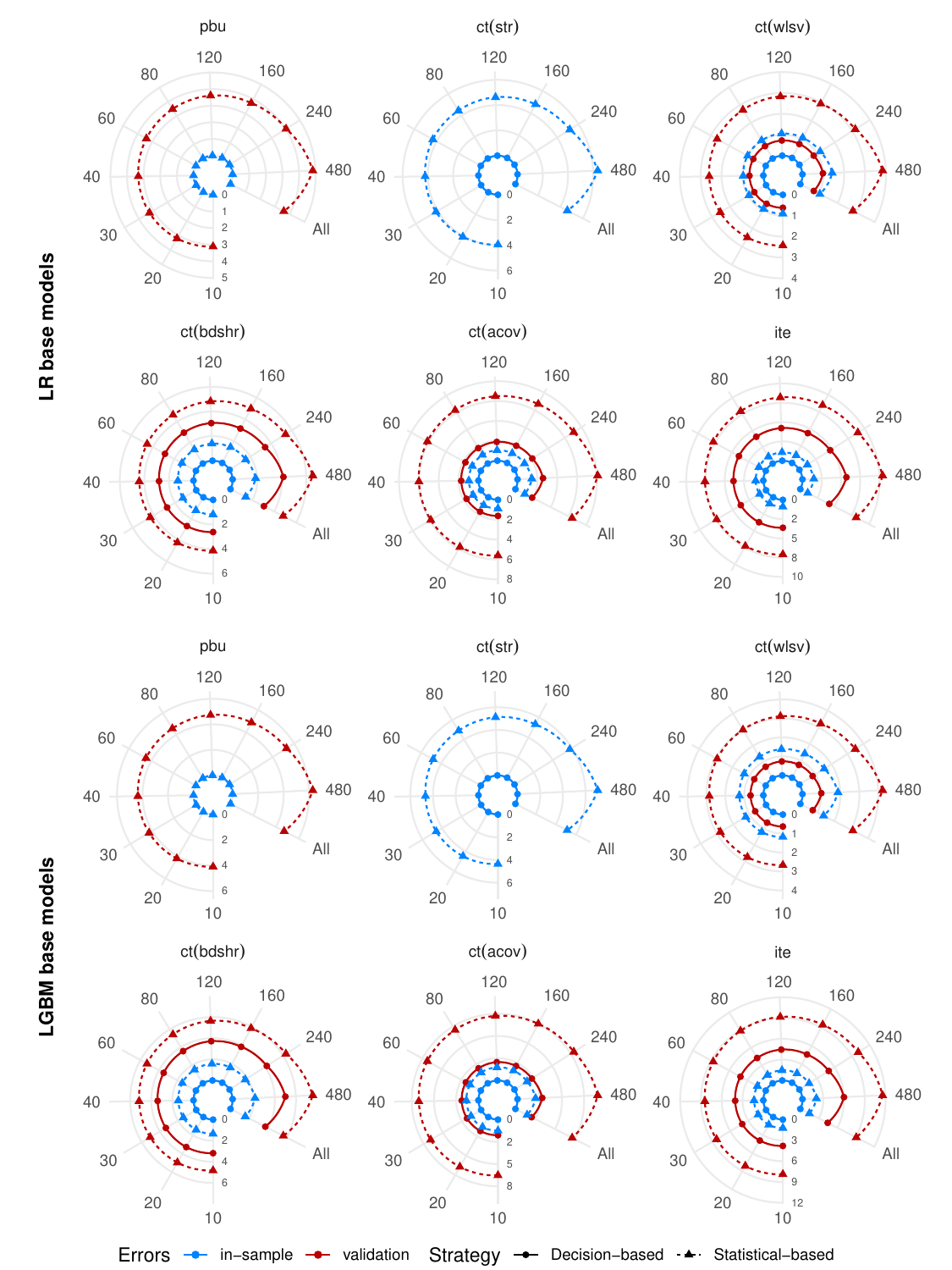}
    \caption{Graphical representation of the improvements in forecast accuracy for different strategies (decision-based \textit{vs} statistical-based) and error types (in-sample \textit{vs} validation) when estimating the covariance matrix. The base models -- LR and LGBM -- and the reconciliation approaches -- pbu, $\text{ct}(str)$, $\text{ct}(wlsv)$, $\text{ct}(bdshr)$, $\text{ct}(acov)$, and ite -- are fixed. For the temporal aggregation $k\in\{10,20,30,40,60,80,120,160,240,480,\text{All}\}$: a closer point to the center indicates greater improvement in forecast accuracy, while points closer to the edge of the circle represent smaller improvements.}
    \label{fig:polar-se}
\end{figure}

To further evaluate the relative performance of these approaches, \autoref{fig:polar-se} presents the percentage improvements in terms of AvgRelMSE, standardized by the worst scenario, i.e., the in-sample errors with a decision-based strategy:
$$
100\left(1-\frac{AvgRelMSE_{j_{e,s}}^{[k]}}{AvgRelMSE_{j_{\text{in},\text{DB}}}^{[k]}}\right)
$$
with $j$ indicates reconciliation approach, $s \in \{\text{DB}, \text{SB}\}$ and $e \in \{\text{in}, \text{vl}\}$.
The results are represented with red and blue for the validation and in-sample errors, respectively, while the continuous and dashed lines distinguish between decision- and statistical-based strategies. For example, when examining the iterative approach with LGBM base forecasts (shown in the last panel), using validation errors improves more than 7\% across different temporal levels, confirming the advantage of incorporating temporal dependencies and out-of-sample set. These graphical insights align with the numerical results and statistical tests, showing the better ability of validation errors when employing a complete temporal structure.

In conclusion, two key findings emerge from this evaluation. The first is that the use of more temporal aggregation levels significantly improves forecast accuracy. By incorporating additional temporal aggregation levels, reconciliation methods are able to capture dynamics that are crucial for improving forecast accuracy. This finding underlies the importance of considering not just short-term forecast horizons, but also longer horizons and various aggregation levels.
The second key finding is that the use of in-sample errors is insufficient to accurately capture the full temporal dynamics of the forecasting problem. Although in-sample errors provide a measure of how well a model fits the data it has already seen, they do not fully account for the out-of-sample forecasting performance, which is critical for the cross-temporal reconciliation. 
However, the use of an out-of-sample validation set ensures a more robust estimation of the cross-temporal covariance matrix, providing a more accurate evaluation of the model’s performance and better capture the complex relationships inherent in temporal data. 

\subsection{Decision cost}

In this study, we introduce two key performance metrics aimed at assessing the economic impact of forecasts in the energy production sector, particularly in relation to regulatory compliance and potential revenue loss. Energy companies are required to submit energy production forecasts to market regulators, specifying the amount of energy they intend to put into the market over the next eight hours. Deviations from these forecasts can lead to financial consequences: if production falls short of the declared (forecasted) amount, companies expose themselves to fines and penalties due to regulatory costs; if production exceeds the forecast, companies have unrealized profits since the excess energy typically cannot be used. We call these two metrics \textit{fines and penalties} ($\delta^{-}$) and \textit{revenue loss} ($\delta^{+}$), respectively, capturing the profit reduction or potential revenue loss associated with production forecast errors.
\begin{itemize}[leftmargin = 2em]
    \item \textit{Fines and penalties} ($\delta^{-}$)\\
    The \textit{fines and penalties} index, denoted by $\delta^{-}$, represents the proportion of total profit for an eight-hour production period that is lost due to fines and regulatory penalties when actual energy production falls below the forecast amount. In other words, when a company forecasts an energy output of $\bar{y}$, but produces only $y < \bar{y}$, the resulting shortfall ($\bar{y} - y$) forces market regulators to find the missing energy from other providers at a higher cost. Consequently, the market regulator imposes fines and penalties on the underperforming company to cover these higher costs. Therefore, $\delta^{-}$ measures the percentage of expected profit consumed by fines and penalties when actual production does not meet the forecast levels.
\item \textit{Revenue loss} ($\delta^{+}$)\\
The \textit{revenue loss} index, denoted by $\delta^{+}$, is a measure of the percentage of additional income that could have been earned. When a company forecasts an output of $\bar{y}$, but the actual production exceeds this amount with $y > \bar{y}$, excess production ($y - \bar{y}$) cannot typically be put on the market due to the regulatory and consumption limits, resulting in unrealized profit. Thus, $\delta^{+}$ quantifies the economic loss associated with overproduction.
\end{itemize}

The indices $\delta^{-}$ and $\delta^{+}$ are calculated by comparing the difference between the forecasted production using the approach $j$ $\left(\bar{y} = \bar{y}_{q, i, j}^{[k]}\right)$ and the actual production $\left(y = y_{q, i}^{[k]}\right)$ relative to the minimum of these two values for the time series 'Total' in temporal aggregation order $k \in \{10, 60, 480\}$ (10 mins, 1 hour and 8 hours), and the forecast origin $q = 1, \dots, Q$. Mathematically, this is represented as follows:
\begin{equation}\label{eq:indices}
    \begin{cases}
    e_{q,j,k}^{-} = \displaystyle\frac{\bar{y}_{q, 1,j}^{[k]}-y_{q, 1}^{[k]}}{y_{q, 1}^{[k]}} & \text{if } y_{q, 1}^{[k]} < (1-\Delta)\bar{y}_{q, 1, j}^{[k]}\\[1em]
    e_{q,j,k}^{+} = \displaystyle\frac{y_{q, 1}^{[k]}-\bar{y}_{q, 1, j}^{[k]}}{\bar{y}_{q, 1, j}^{[k]}} & \text{if } y_{q, 1}^{[k]} > (1+\Delta)\bar{y}_{q, 1, j}^{[k]}
\end{cases} 
\end{equation}
where $\Delta$ represents a threshold set to $1\%$. The conclusions and interpretations of the final results remain consistent even if this threshold value is modified (see the Appendix \ref{app:appA}).
This proportion is then averaged across all the forecast origin where underproduction or overproduction occurred, respectivelly:
$$
\delta^{-}_{j,k} = \bigg(\prod_{q \in \mathcal{Q}_{j,k}^{-}} e_{q,j,k}^{-} \bigg)^{\displaystyle\left|\mathcal{Q}_{j,k}^{-}\right|^{-1}} \quad \text{and} \quad \delta^{+}_{j,k} = \bigg(\prod_{q \in \mathcal{Q}_{j,k}^{+}} e_{q,j,k}^{+} \bigg)^{\displaystyle\left|\mathcal{Q}_{j,k}^{+}\right|^{-1}}
$$
where $\mathcal{Q}_{j,k}^{-} = \{q = 1,\dots, Q \; \mid \; y_{q, 1}^{[k]} < \bar{y}_{q, 1, j}^{[k]} \}$ and $\mathcal{Q}_{j,k}^{+} = \{q = 1,\dots, Q \; \mid \; y_{q, 1}^{[k]} > \bar{y}_{q, 1, j}^{[k]}\}$. 

These two indices provide critical insight into the financial/economic implications of forecast accuracy in the context of regulatory compliance and profit maximization for energy producers. Therefore, $\delta^{-}$ and $\delta^{+}$ offer a comprehensive view of how forecast accuracy impacts both regulatory cost management and profit generation, highlighting the importance of precise forecasting in both operational and financial planning for energy companies.

In Appendix \ref{app:appA}, we also report the percentage number of overproduction and underproduction that occurs. It should be noted that there is not a large difference between the approaches, with overproduction occurring in 50 to 57\% of cases, while underproduction is observed in 40 to 47\%, at any temporal aggregation levels. The slightly higher frequency of overproduction events may suggest a conservative tendency to avoid fines and penalties. 

\begin{figure}[!th]
    \centering
    \includegraphics[width=0.95\linewidth]{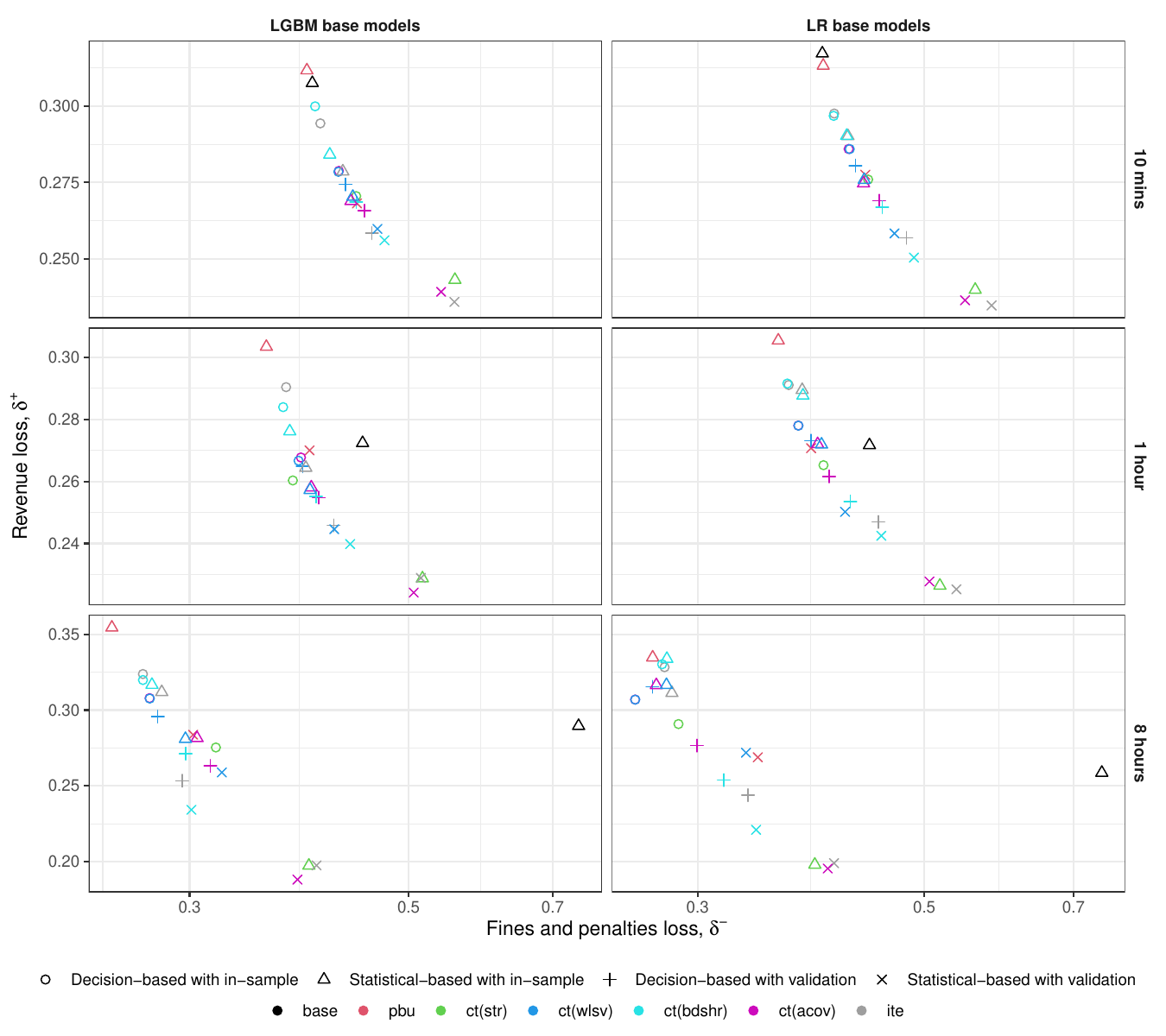}
    \caption{Energy forecast performance metrics for the 10 mins, 1 hour and 8 hours forecasts of the most aggregated time series ($Total$), where $\delta^{-}$ (x-axis) represents the percentage of profit lost due to fines and penalties from underproduction, and $\delta^{+}$ (y-axis) indicates the percentage of unrealized profit resulting from overproduction. Each point corresponds to a different forecasting approach.}
    \label{fig:index}
\end{figure}

\autoref{fig:index} shows the different values of $\delta^{-}$ (x-axis) and $\delta^{+}$ (y-axis) for the 10 mins (first row), 1 hour (second row) and 8 hours (third row) forecasts of the most aggregated time series ($Total$). At the 10 mins, the base and pbu approaches are highly conservative, achieving the smallest $\delta^{-}$ but at the cost of the highest $\delta^{+}$. In contrast, methods like ite and ct(acov) are less conservative with smaller revenue losses. When considering 1 hour aggregated time series, the base approach loses its effectiveness in minimizing $\delta^{-}$, with other methods starting to outperform them. Finally, at the 8 hours aggregation level, the base approach tends to overestimate forecasts, reducing revenue loss but showing higher fines and penalties losses. In general, the results suggest that the statistically based hierarchy tends to adopt less conservative forecasts regarding $\delta^{-}$ and reduce revenue losses. In contrast, decision-based methods offer a more balanced compromise between $\delta^{-}$ and $\delta^{+}$. Specifically, decision-based strategies with validation errors, such as $ct(bdshr)$ and $ct(wlsv)$, effectively balance the accuracy of the forecast with the economic performance, highlighting their versatility and effectiveness on different temporal aggregation levels.

\clearpage
\section{Conclusions}\label{sec:conclusion}
In this study, we explored advanced cross-temporal forecasting models and their potential to enhance wind power forecasting accuracy. By introducing validation errors for covariance matrix estimation, we proposed a robust alternative to traditional in-sample error-based approaches in hierarchical forecasting, addressing the challenges posed by the stochastic and intermittent nature of wind power.

We developed cross-temporal hierarchies inspired by practical decision-making, focusing on temporal granularities relevant to operational decisions, and compared them with statistical-based cross-temporal hierarchies. Our empirical analysis revealed that while statistical-based approaches achieve higher accuracy, decision-driven models offer a practical balance between accuracy and economic performance, particularly under varying temporal frequencies.

Additionally, we evaluated the forecasting performance of the models not only based on accuracy but also on the costs they incur for wind farms. This integration of decision costs into model evaluation underscores the importance of aligning forecasting objectives with operational priorities in the energy market. Our findings highlight that decision-based methods, particularly those leveraging validation error strategies such as ct(bdshr) and ct(wlsv), provide a versatile and effective framework for balancing forecast accuracy with real-world economic impacts. Specifically, statistical-based hierarchies tended to adopt less conservative forecasts, reducing revenue losses. In contrast, decision-based methods offered a more balanced compromise between accuracy and decision costs, making them particularly attractive for practical applications.

These results contribute to advancing the fields of wind power forecasting and hierarchical forecasting models, demonstrating that such models can significantly enhance forecasting accuracy and, consequently, the reliability and profitability of renewable energy integration into the grid. Future research could extend these insights by exploring machine learning models for non-linear reconciliation of time series. Another promising direction is the application of multi-objective optimization methods to balance or globally optimize forecast accuracy and decision costs, further improving the practical efficacy of forecasting models in real-world applications.


  \printbibliography

\clearpage
\appendix

\section{Extended figures for operational evaluation}\label{app:appA}

This section provides an extended analysis of operational performance, focusing on the relationship between forecast accuracy and its implications for both operational reliability and economic outcomes. The figures included offer detailed insights into the effects of forecast errors across varying conditions and performance metrics.

\begin{figure}[h]
    \centering
    \includegraphics[width=0.8\linewidth]{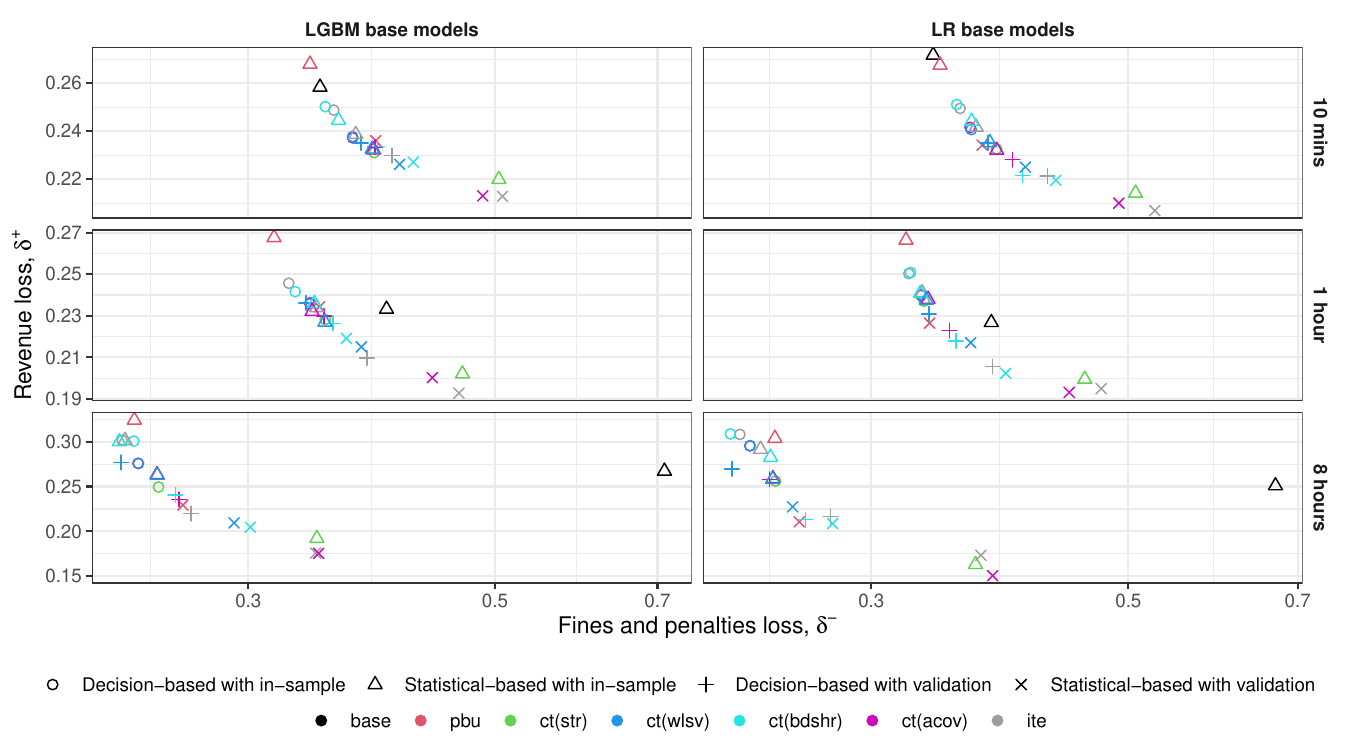}
    \caption{Energy forecast performance metrics ($\Delta = 0$) for the 10 mins, 1 hour and 8 hours forecasts of the most aggregated time series ($Total$).}
    \label{fig:index_app1}
    \includegraphics[width=0.8\linewidth]{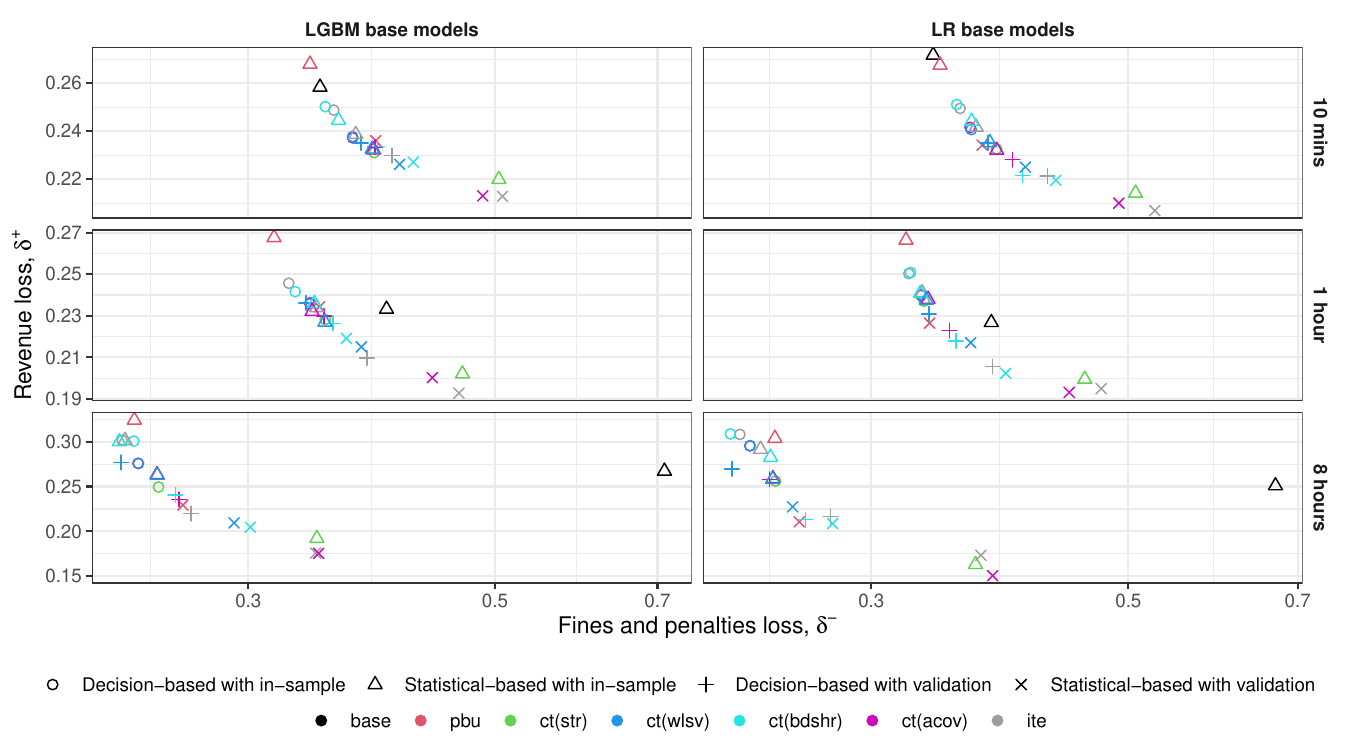}
    \caption{Energy forecast performance metrics ($\Delta = 2.5\%$) for the 10 mins, 1 hour and 8 hours forecasts of the most aggregated time series ($Total$).}
    \label{fig:index_app2}
\end{figure}

\begin{figure}[h]
    \centering
    \includegraphics[width=0.9\linewidth]{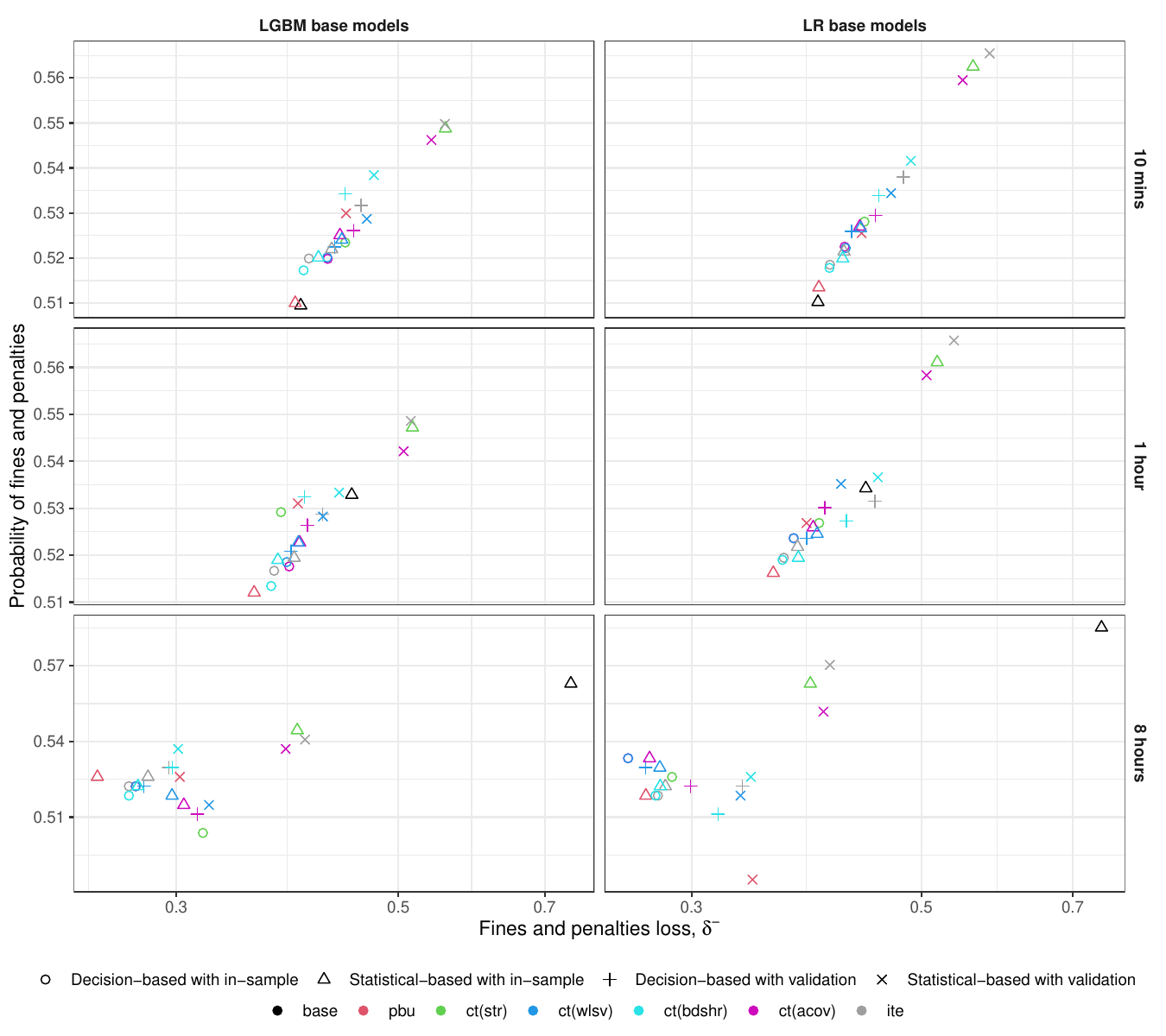}
    \caption{Energy forecast performance metrics for the 10 mins, 1 hour and 8 hours forecasts of the most aggregated time series ($Total$), where $\delta^{-}$ (x-axis) represents the percentage of profit lost due to fines and penalties from underproduction, while the y-axis shows the percentage of time underproduction occurs.}
    \label{fig:index_app3}
\end{figure}

\begin{figure}[h]
    \centering
    \includegraphics[width=0.9\linewidth]{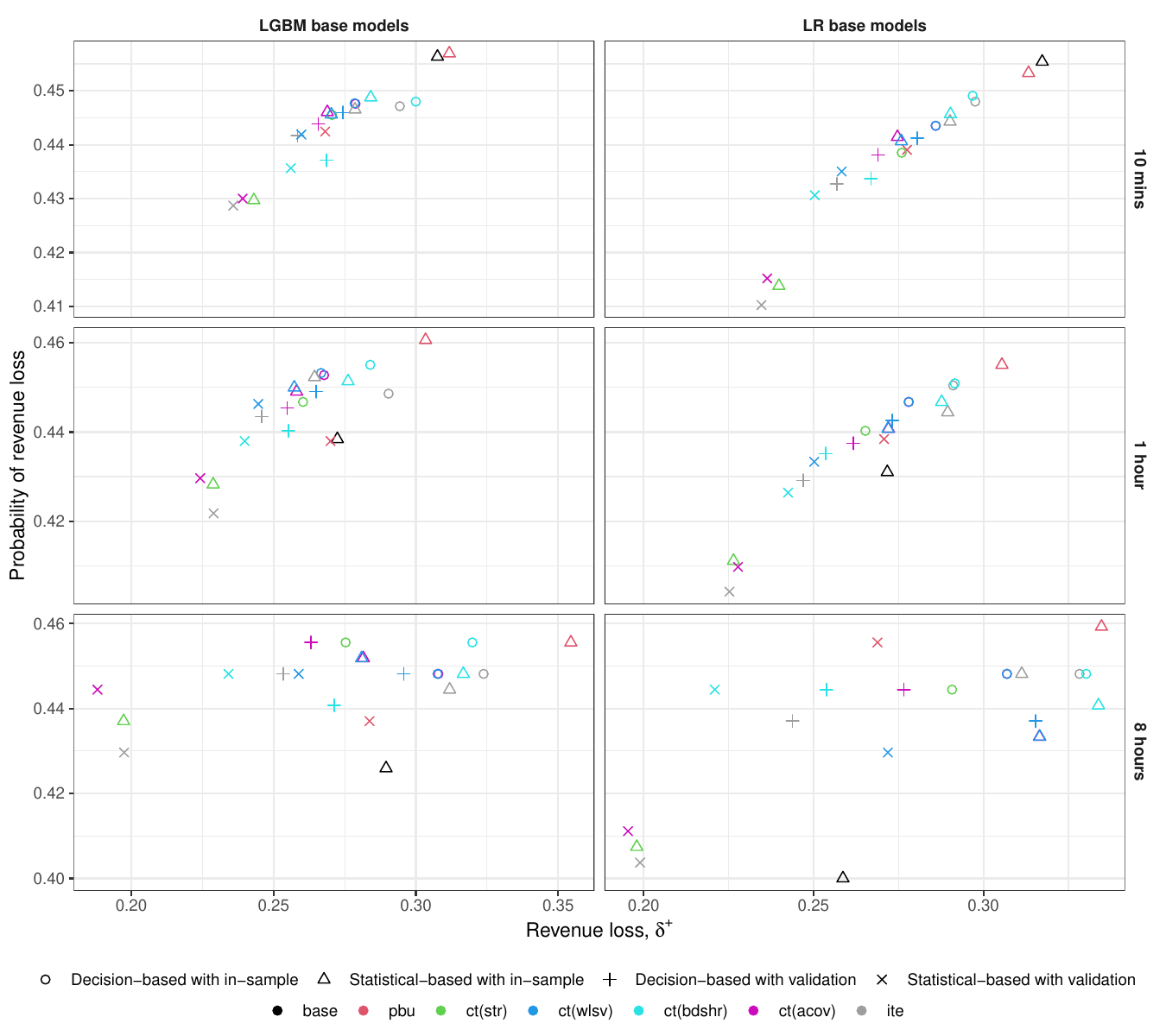}
    \caption{Energy forecast performance metrics for the 10 mins, 1 hour and 8 hours forecasts of the most aggregated time series ($Total$), where $\delta^{+}$ (y-axis) represents the percentage of unrealized profit resulting from overproduction, while the y-axis indicates the percentage of time overproduction occurs.}
    \label{fig:index_app4}
\end{figure}

Figures \ref{fig:index_app1} and \ref{fig:index_app2} represent the indices $\delta^+$ and $\delta^-$ under different threshold values of $\Delta$, as defined in Equation~\ref{eq:indices}. Figure~\ref{fig:index_app3} explores the impact of underproduction, illustrating the percentage of profit lost due to fines and penalties ($\delta^-$) along with the percentage of time such underproduction occurs. In contrast, Figure~\ref{fig:index_app4} examines the implications of overproduction, presenting the percentage of unrealized profit ($\delta^+$) and the frequency of overproduction events.

\end{document}